\documentclass[reprint,
superscriptaddress,
frontmatterverbose, 
preprintnumbers,
nofootinbib,
nobibnotes,
amsmath,amssymb,
aps,
prx,
floatfix,
]{revtex4-2}

\usepackage{graphicx}
\usepackage{dcolumn}
\usepackage{bm}
\usepackage[mathlines]{lineno}
\usepackage[colorlinks=true,urlcolor=blue,linkcolor=blue,citecolor=blue]{hyperref}
\usepackage{color}
\usepackage{xcolor}
\begin{document}
		\title{Synergistic approach towards reproducible high zT in superionic thermoelectric Ag$_2$Te}
	\author{Navita Jakhar}
	\affiliation{Department of Physics, Indian Institute of Science Education and Research, Dr. Homi Bhabha Road, Pune 411 008, India}
	\author{Nita Bisht}
	\affiliation{Savitribai Phule Pune University Ganeshkhind Pune 411007, India}
	\author{Ankita Katre}
	\affiliation{Savitribai Phule Pune University Ganeshkhind Pune 411007, India}
	\author{Surjeet Singh}
	\email{surjeet.singh@iiserpune.ac.in}
	\affiliation{Department of Physics, Indian Institute of Science Education and Research, Dr. Homi Bhabha Road, Pune 411 008, India}
	
	\date{\today}
		\begin{abstract}
		
		Recently, the superionic thermoelectrics, which typify the novel `phonon-liquid electron-crystal'\  concept, have attracted enormous attention due to their ultralow thermal conductivity and high figure-of-merit (zT). However, their high zT is generally obtained deep inside the superionic phase, e.g., near 1000~K in Cu$_2$X (X: chalcogen atom) family where the superionic transition is close to 400~K. At such high temperatures, the liquid-like flow of the metal ions under an electric field or a temperature gradient, both of which are integral to the working of a thermoelectric device, results in device degradation. To harness the full potential of the superionic thermoelectrics, it is therefore necessary to reach high zT at low temperatures where the metal-ion diffusion is not an issue. Here, we present a novel all-room-temperature route to fabricate 100\% dense, nanostructured Ag$_2$Te with highly reproducible thermoelectric properties and a high zTof 1.2 at 570~K, i.e., merely 150~K above its superionic transition. The samples show a broad particle-size distribution ranging from a few nm to a few $\mu$m. This hierarchical nanostructuring is shown to suppress the thermal conductivity of Ag$_2$Te beyond the phonon-liquid electron-crystal limit to ultra low values, leading to a remarkable enhancement of 87\% in the zT over that of the ingot sample. These values supersede the zT of any Ag$_2$Te previously reported. Our results are supported by first-principles density functional theory calculations of the electronic and thermal properties.\\	
	\end{abstract}

\maketitle

\section{Introduction}
\label{Intro}
	
Mankind's overdependence on fossil fuels to meet the constantly growing energy demand worldwide has led to climate change and global warming. For a sustainable future, it is important to develop alternative means of producing clean energy. Using thermoelectric devices, waste heat can be converted directly into clean and useful electrical energy; alternatively, these devices, with some design modification, can also be employed for producing environmentally friendly air conditioning and refrigeration modules. The conversion efficiency of a thermoelectric devices is governed by the dimensionless figure-of-merit $\rm zT$ of the n-type and p-type thermoelectric pairs used in their fabrication~\cite{bell2008cooling,he2017advances}. The $\rm zT$ at any temperature T (absolute) is given by $\rm zT = (\sigma S^2/\kappa)T$ where S is the Seebeck coefficient and $\sigma$ is the electrical conductivity. Both these quantities are closely, and generally inversely, related. Their combination in the form $\rm S^2\sigma$ is known as the power factor (PF). The quantity $\kappa$ is the thermal conductivity of the material, which can be expressed as an algebraic sum of the lattice thermal conductivity ($\kappa_l$) and the electronic thermal conductivity ($\kappa_e$). 
Due to their contraindicated interrelationship, decoupling $\sigma$, $\rm S$ and $\kappa$ in a material is generally non-trivial, making the task of enhancing  zT highly challenging. However, despite this challenge, the zT of bulk thermoelectric materials has shown remarkable improvement over the past two decades. This can be attributed to the development of several novel concepts which can either decouple $\sigma$ and $\rm S$ to enhance the PF or suppress the lattice component of the thermal conductivity significantly without adversely affecting  the PF~\cite{sootsman2009new}. Some of the important examples of these approaches include: charge-carrier filtering~\cite{zide2006demonstration, faleev2008theory}, resonant doping~\cite{heremans2008enhancement, wu2017resonant, liu2012convergence, pei2011convergence}, band convergence~\cite{heremans2008enhancement,  wu2017resonant, liu2012convergence, pei2011convergence}, nanostructuring~\cite{dresselhaus2009kohn, vineis2010nanostructured, kanatzidis2010nanostructured}, all-scale hierarchical phonon scattering~\cite{biswas2012high, wang2011thermal, selli2016hierarchical}, optimizing atomic ordering~\cite{roychowdhury2021enhanced}, or by the phonon-glass electron-crystal (PGEC) approach~\cite{rowe2018crc}. In particular, in the PGEC class of thermoelectrics, the electronic transport properties are typical of crystalline semiconductors but the lattice thermal conductivity is characteristic of glasses or amorphous materials, resulting in intrinsically low values of $\kappa_l$ as in the Zintl phase compounds \cite{toberer2010zintl}, filled skutterudites \cite{shi2011multiple}, and misfit-layered cobalt oxides \cite{miyazaki2004crystal}, among others. Recently, a related but microscopically different concept of phonon-liquid electron-crystal (PLEC) has been proposed to explain the ultralow thermal conductivity of compounds Cu$_2$X (X = S, Se, and Te) in their superionic phase leading to high zT values. For example, a high zT value of 1.1 at 1000~K in Cu$_2$Te~\cite{he2015high}, 1.7 at 1000~K in Cu$_{2-x}$S~\cite{he2014high}, and 2.3 at 1000~K in Cu$_{2-x}$Se~\cite{liu2013ultrahigh} are recently reported. In the superionic phase above $\rm T_t \sim 400~K$, the Cu ions are in liquid-like phase. At temperatures much higher than $\rm T_t$, such as 1000~K where the highest zT values are reported for these materials, the Cu-ions tend to migrate under a temperature gradient or a moderate applied electric field, both intrinsic to the operation of a thermoelectric device, leading to sample deterioration or decomposition~\cite{Brown2013, dennler2014binary}. In the last few years, innovative methods are being employed to overcome this issue in Cu$_2$X superionics~\cite{TaoMao2020, Dongwang2020, Kunpeng2022}. 

Here, we investigate Ag$_2$Te which belongs to a technologically important class of semiconducting/superionic materials Ag$_2$X (X = S, Se, or Te). Upon heating, Ag$_2$Te undergoes a superionic  transition near $\rm T_t = 420$~K. 
In the  room-temperature phase, it has an intrinsically low thermal conductivity of $\rm \sim 1.2~W~m^{-1}~K^{-1}$ which decreases further to values as small as $\rm \sim0.7~W~m^{-1}~K^{-1}$ in the superionic phase ~\cite{capps2010significant, zhu2015enhanced}. 
The carrier mobility in Ag$_2$Te is high $\rm \sim 10^3~cm^2~V^{-1}s^{-1}$~\cite{capps2010significant}, and the carriers mobility ratio ($\rm \mu_e$/$\mu_h$) 
is around 6~\cite{taylor1961thermoelectric}. The Seebeck coefficient at room-temperature is negative and has a reasonably high value of $\rm -100~\mu V~K^{-1}$ 
despite minority carrier excitations due to a small bandgap $\rm E_g$ of  $< 0.1$~eV~\cite{jahangirli2018ab,capps2010significant}. A very low-thermal conductivity combined with high carrier mobility and high thermopower makes Ag$_2$Te a promising thermoelectric material. In Ag$_2$Te ingots prepared by solid state melting, a zT of 0.64 at 575~K was previously reported~\cite{capps2010significant}. 
This was further enhanced to 1 at 575~K by PbTe doping through bandgap engineering~\cite{pei2011alloying}. 
These studies suggest that in Ag$_2$Te based materials, through further optimization, high zT values can be achieved at temperatures not too high above the superionic transition. This makes tackling the metal-ion migration issue in the superionic phase relatively easier in the Ag compound than in its Cu counterpart where a high zT is attained only near 1000~K (the zT around 600~K in all cases being very low)~\cite{bailey2017potential}.   

However, despite these remarkable properties Ag$_2$Te has remained at the backseat in this class of materials. The dark side of Ag$_2$Te relates to its extreme sample dependence. While some reports show Ag$_2$Te to undergo a sharp transition in thermopower from negative (n) to positive (p) across $\rm T_t$ upon heating~\cite{cadavid2013organic,aliev2003electrical, dalven1966energy, wood1961degeneracy}, in some other reports the thermopower is shown to remain negative, displaying a step-like increase at $\rm T_t$~\cite{pei2011alloying,cadavid2013organic,chang2019facile,capps2010significant,taylor1961thermoelectric,jung2012effect,fujikane2005thermoelectric,wood1961degeneracy}.  
It is believed that the n-type or p-type behavior in the superionic phase arises due to very small Ag off-stoichiometry: n-type for a slight Ag excess and p-type for Ag deficient samples. Interestingly, within the n-type samples also, the electrical conductivity at room temperature exhibits a significant dispersion varying from 2650~S~cm$^{-1}$ in Ag$_2$Te ingot samples~\cite{aliev2015dependence} to 800~S~cm$^{-1}$ or even less in hot pressed chemically synthesized Ag$_2$Te nanopowders~\cite{cadavid2013organic}.   It is therefore evident that the thermoelectric properties of Ag$_2$Te are extremely sensitive to the exact Ag/Te molar ratio, which varies unpredictably depending on the magnitude of temperature gradient present during the sintering process. This significant sample dependence prevents further improvement in zT using various strategies mentioned earlier. 

We first mitigate this issue by developing a novel all-room-temperature method of fabricating highly reproducible and 100\% dense Ag$_2$Te samples. We then go beyond the PLEC limit by adopting a hierarchical nanostructuring approach spanning length scales ranging from few nm to $\mu$m. This is achieved by using a low frequency vibratory ball mill which produces a broad particle-size distribution, a rich variety of particle shapes and a varying degree of crystallinity with dislocations present over length scale as small as few nm. The ball milled nanopowders were directly cold-pressed to achieve densities close to 100\% of the theoretical density, which completely precluded the need for high-temperature sintering or spark-plasma sintering used in the previous works to reinforce high-density or nanostructuring. The realization that Ag$_2$X materials are ductile in nature played a crucial role in this innovation~\cite{shi2018room}. Further, by increasing the milling duration, the lattice thermal conductivity is systematically reduced.  We thus obtain a very high zT of 1.2 at 570~K in our n-type samples and 0.64 at 570~K in the p-type samples. To appreciate why these number are very high in the context of Ag$_2$Te, in Fig.~\ref{XRD}d we have compared the zT of our samples with the highest zT reported for Ag$_2$Te in various previous studies. The zT values in our study not only supersede the best zT for the undoped samples but also for the doped samples. The highest \textit{average} zT of 0.99 in our n-type Ag$_2$Te is also comparable to that of doped Bi$_2$Te$_3$, in the temperature range of 300~K to 570~K~\cite{malik2020enhanced}. 
The density functional theory (DFT) based electronic  structure and thermal transport calculations are also performed to understand the n-type to p-type transition, and the origin of  intrinsically low thermal conductivity and its further suppression by nanostructuring. 
 \section{RESULTS AND DISCUSSION}
\subsection{Nomenclature}
Several Ag$_2$Te samples were prepared during this study using various methods, including (a) solid-state melting (SSM), (b) Ball-milling followed by cold-pressing (BMC), (c) Ball-milling followed by cold-pressing and high-temperature sintering (BMS), (d) Hand-milling followed by cold-pressing and high-temperature sintering (HMS). For the ease of reading, the samples are labeled as AGs-tttXYZ where XYZ = \{SSM, BMC, BMS, HMS\}; ttt refers to the milling/grinding time in minutes for the ball-milled/hand-grind samples. The quantity s in ‘AGs’ has values s = \{1, 2, 3,...\}, designating the sample number within a XYZ series. For Ag excess or deficient samples, the same nomenclature has been used but with an extension ‘$\rm +x$’ or `$\rm-x$/' where x depicts the percentage of Ag excess ($+$) or Ag deficiency ($-$). The complete details of sample preparation and a list of all the samples reported here is given in the experimental section. 

\subsection{Structural Characterization}
\begin{figure*}[!]
	\includegraphics[width=16cm]{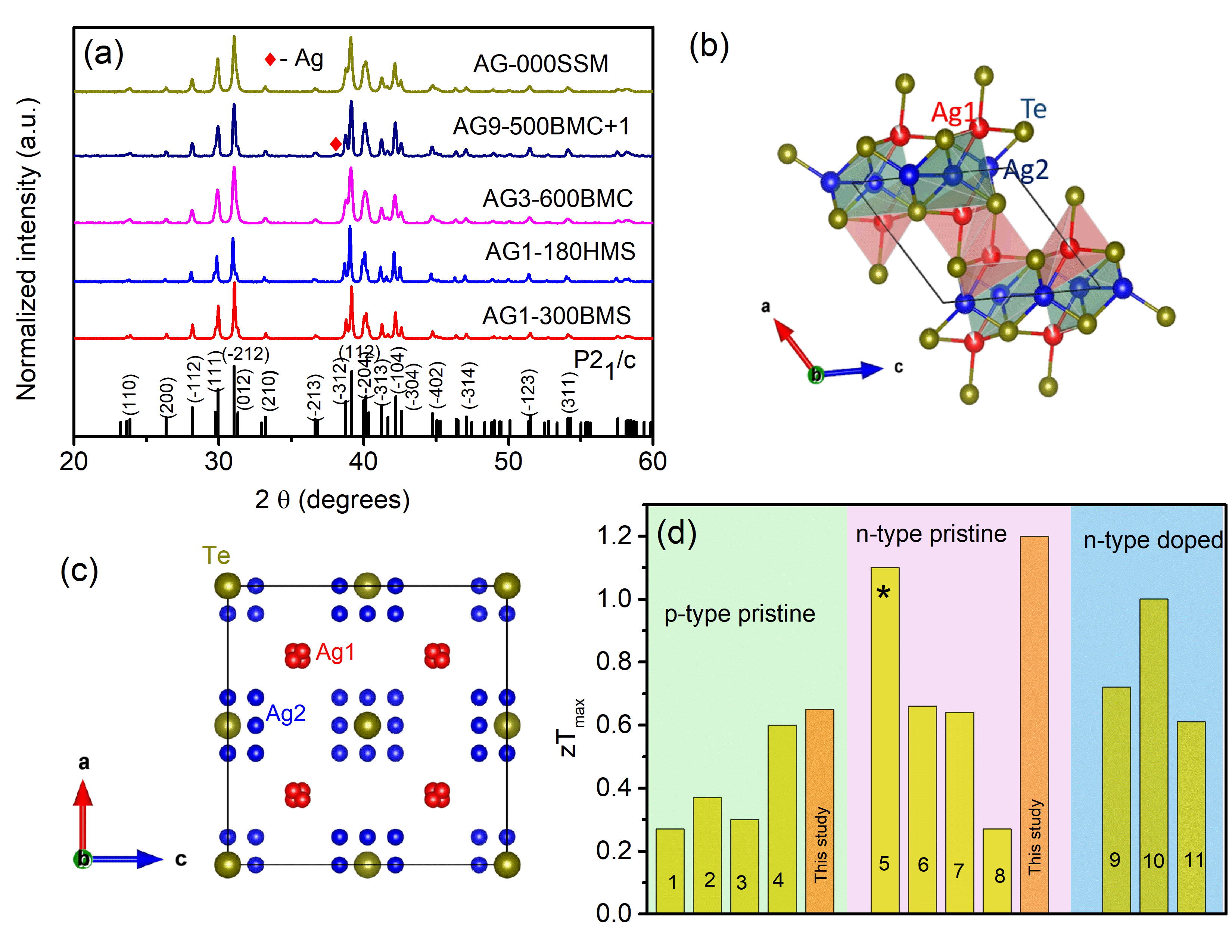}
	\caption{(a) X-ray powder diffraction patterns for some representative Ag$_2$Te samples, (b) Crystal structure of room-temperature monoclinic $\beta$-Ag$_2$Te, and (c) Crystal structure of the high temperature cubic $\alpha$-phase; (d) The maximum zT in our study is compared with that reported in various previous works for p-type pristine, n-type pristine and n-type doped Ag$_2$Te samples. (1) Cadavid et al.\cite{cadavid2013organic}, (2) Yang et al.\cite{yang2014composition}, (3) Cadavid et al. \cite{cadavid2012bottom}, (4) Zhu et al. \cite{zhu2015enhanced}, (5) Chang et al.\cite{chang2019facile}. * sample density $\approx$~70\%, which implies that the actual zT will be much smaller than the reported value,  (6) Cadavid et al.\cite{cadavid2013organic}, (7) Capps et al.\cite{capps2010significant}, (8) Fujikane et al.\cite{he2015high}, (9) Hu et al.\cite{hu2021fast}, (10) Pie et al.\cite{pei2011alloying}, and (11) Zhou et al.\cite{zhou2012preparation}}
	\label{XRD}
\end{figure*}

The phase purity of our samples was checked using the x-ray powder diffraction (XRPD) method at room-temperature. A few representative results are shown in~Fig.~\ref{XRD}a. The observed XRPD patterns can be satisfactorily indexed on the basis of the monoclinic ($\rm P2_1/c$) symmetry with no extra peaks. The only exception to this are the samples prepared with excess Ag where a small extra peak due to unreacted Ag metal is also seen as shown for a representative sample in Fig.~\ref{XRD}a. The presence of structural phase transition at $\rm T_t$ was confirmed using the Differential Scanning Calorimetry shown in the Supporting Material, Fig.~S1. The cubic $\rm Fm\bar{3}m$ symmetry in the superionic phase was confirmed using the high-temperature XRPD Fig.~S2. Hereafter, we shall refer to the monoclinic  phase below $\rm T_t$ as $\beta$-Ag$_2$Te, and the cubic phase above $\rm T_t$ as 
$\alpha$-Ag$_2$Te. The crystal structure of $\beta$-Ag$_2$Te is layered comprising Ag(1) and Ag(2)-Te layers as shown in Fig.~\ref{XRD}b, where Ag(1) and Ag(2) are inequivalent Ag sites having different Te coordination. In the high-temperature phase shown in Fig.~\ref{XRD}c, while the Te ions form a rigid \textit{fcc} network, the Ag ions hop between various interstitial sites shown in the figure. The lattice constants obtained from the Rietveld refinement are given in Table~S1, Supporting Information. A good agreement with the reported lattice parameters is found for all our samples~\cite{van1993redetermination}. 
 
\begin{figure*}[!]
	\includegraphics[width=17cm]{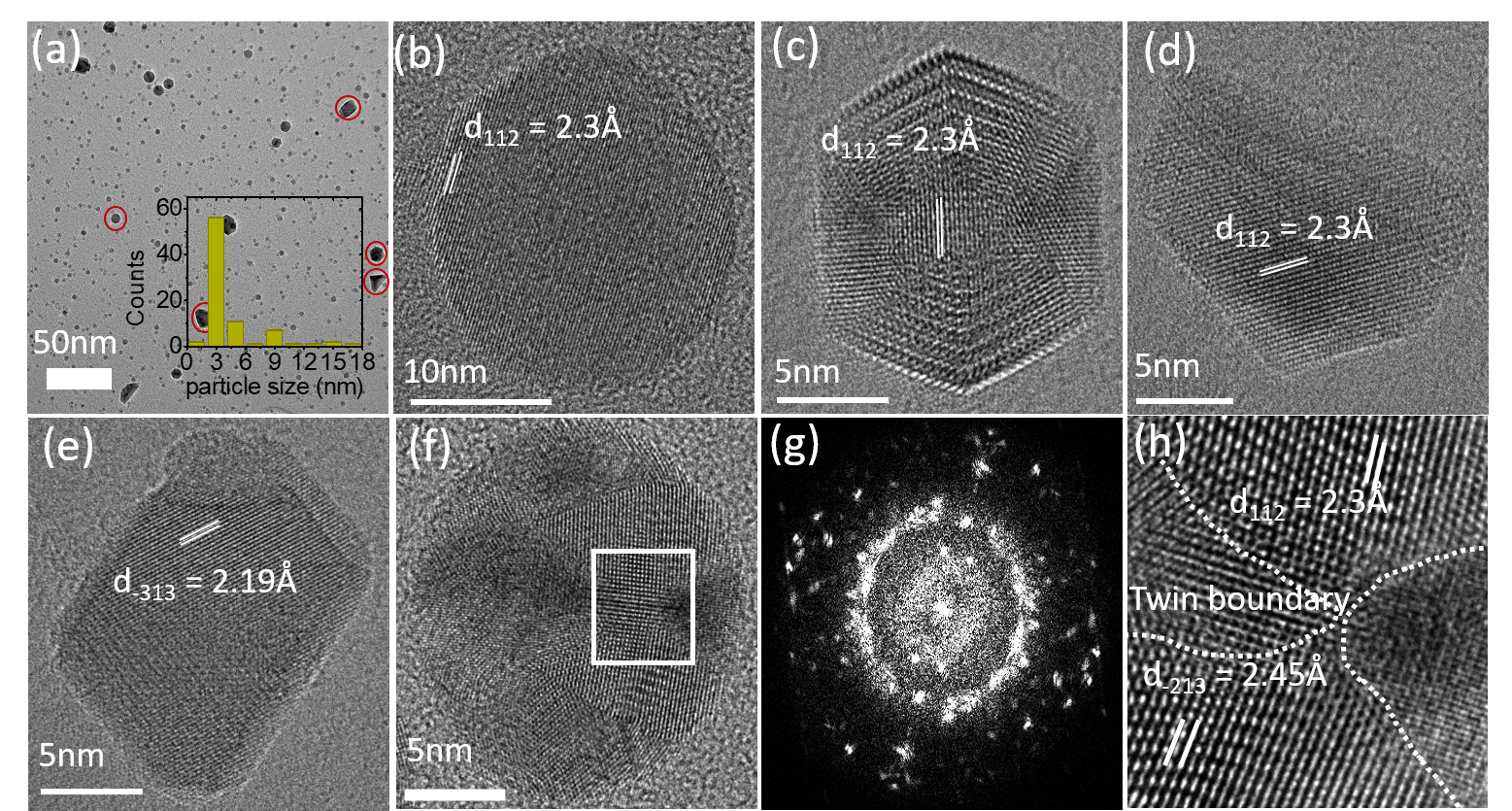}
	\caption{  TEM analysis of the Ball-milled powders (a) a low resolution TEM image showing Ag$_2$Te nanoparticles. Inset shows the particle-size distribution, (b - f) high-resolution TEM images showing different morphologies and degree of crystalline order in the Ag$_2$Te nanoparticles in (a), (g) FFT of nanoparticle in (g), and (h) a zoomed-in image on the bordered area (in white) in (g). The dashed lines showing the grain boundaries are drawn as a guide to eye.}
	\label{TEM}
\end{figure*}

A few representative HRTEM images of the Ag$_2$Te nanoparticles are shown in~Fig.~\ref{TEM}. The HRTEM samples were prepared by centrifuging the ball-milled powder in an ethanol medium. This makes the heavier (and bigger) particles to settle down in the centrifuged vial, allowing only the smaller particles to be picked by the pippet for examination. The image in panel (a) is a low-resolution TEM image showing the particle-size distribution in our samples. The ball-milled nanoparticles show a surprisingly rich diversity in their shapes and the degree of crystallinity. The image in panel (b), for example, shows a spherical particle showing a high degree of crystallinity with nicely lined-up lattice planes. On the other hand, the particle in (c) shows a hexagonal shape with lattice planes near the top and bottom aligned parallel to the edges of the grain; the alignment is horizontal close to the left and right edges, and vertical around the center of the particle. Similarly, in panel (d) a pentagonal shaped particle, and in (e) a rectangular shaped particle with spatially varying crystalline orientation over nanometer length-scale are shown. To demonstrate this diversity further, a high-resolution HRTEM image of a nearly spherical nanoparticle of diameter 20~nm is shown in panel (f). Unlike the spherical particle in (b), the particle in panel (f) shows a spatially varying crystalline order akin to the other less symmetric particles in other panels. A zoomed-in image of the region in the white box in (f) is shown in (h). It shows four differently oriented grains over an area of 5 nm by 5 nm. (g) shows the FFT of particle in (f). A few more corroborative images are shown in Fig.~S3, Supporting Information.

These wide variations in shape and crystallinity arises due to ball-milling. The ball-milling procedure was performed in a low-frequency (120 rpm) vibratory mill where a stainless steel (SS) cylindrical shell, loaded with the charge (Ag and Te powders) and SS balls in an inert atmosphere, oscillates along the arc of a circle, like an inverted pendulum. The balls inside thereupon execute a complex oscillatory motion undergoing various types of collisions between themselves and with the walls of the shell~\cite{yokoyama1996simulation}. In general, when this process runs for a short time, depending on the hardness of the material and the initial particle size, the final fineness varies from 0.1$\mu$m to 1~$\mu$m; however, by a continuous repetitive movement in an optimized charge-to-ball mass ratio and ball diameter, smaller particles with a broad particle-size distribution can be obtained. The repetitive collisions happening at glancing angle generate large shear forces tearing a particle into smaller sized particles, leading to decrease in the average particle-size. 

Since, the method used for preparing the HRTEM grids is designed to exclude the larger sized particles and agglomerates, to obtain a full spectrum of the particle-size distribution in our ball-milled samples, FESEM was done on freshly fractured surfaces of BMC samples. A few representative FESEM images are shown in Fig.~S4(a, c), Supporting Material. The image shows densely packed grains ranging in size from a few nm to $\mu$m. The particle-size distribution is shown in the inset. The effect of sintering at 550~K for 6~h is shown in~Fig. S3(b, d). Upon heating, the microstructure remains intact. The grain boundaries appear to have fused. As the melting temperature of Ag$_2$Te is around 1273~K, this is  likely due to the local diffusion of Ag ions across the grain boundaries in the superionic phase. The HRTEM on the heated samples confirms this where nm-sized randomly oriented grains are seen reminiscent of the ball-milled powder (Fig.~\ref{k and zT}f and Fig.~S5).   

\subsection{Charge Transport properties}
\begin{figure*}[!]
	\includegraphics[width= 17cm]{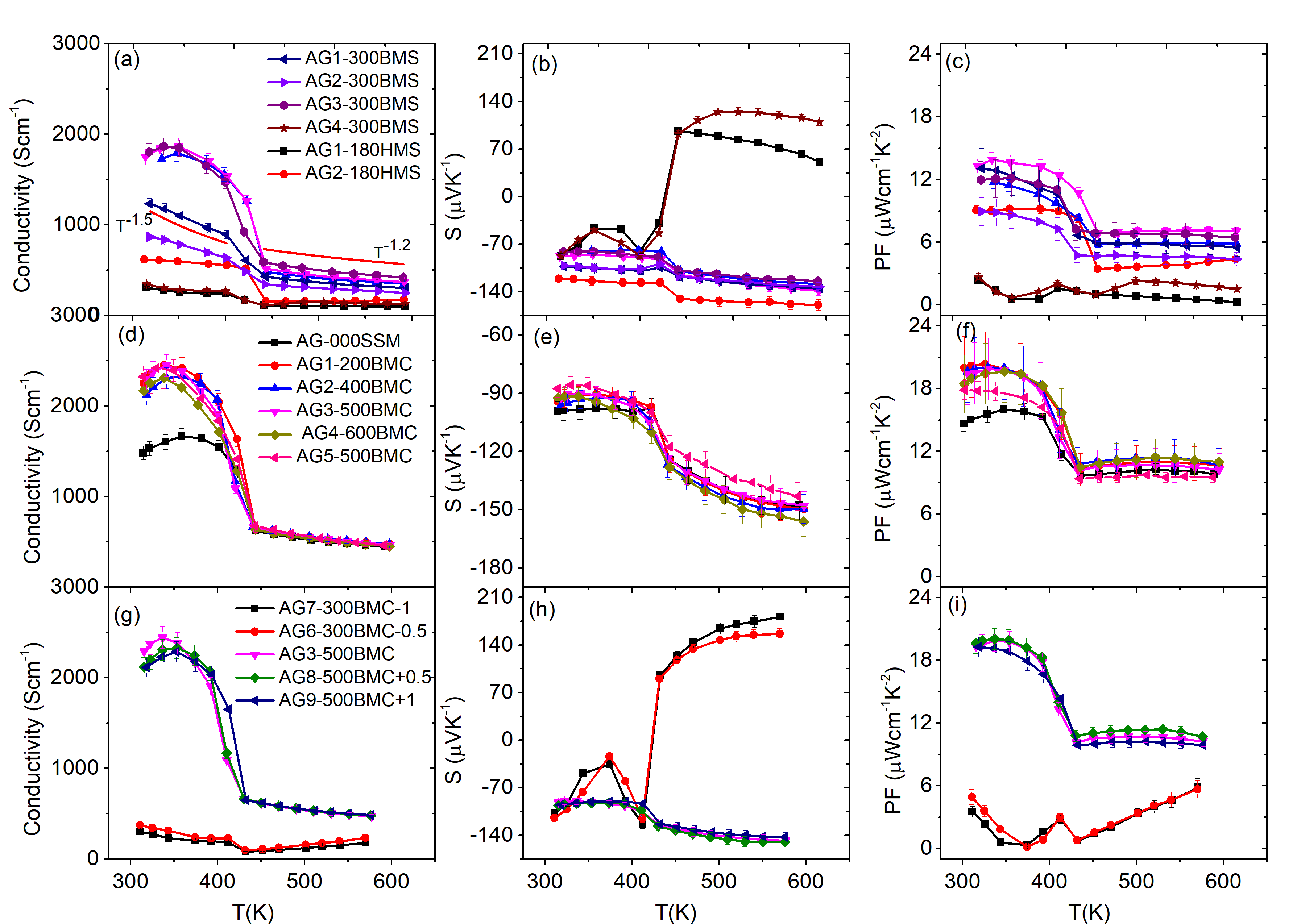}
	\caption{Temperature dependence of electrical conductivity, thermopower, and power factor for our samples. (a - c) BMS and HMS samples, (d - f) Stoichiomteric BMC samples prepared for different milling durations, (g - i) BMC samples prepared with different Ag/Te molar ratio (see text for details).}
	\label{LSR}
\end{figure*}

The temperature dependence of $\sigma$ of our BMS samples, namely, AGs-300BMS (s = 1, 2, 3 and 4) is shown in Fig.~\ref{LSR}a. The sample AG3-300BMS has the highest electrical conductivity of 1805~S~cm$^{-1}$ near 300~K. For this sample, $\sigma$ increases upon heating above 300~K, exhibits a maximum around 330~K and drops precipitously thereafter due to the onset of the superionic transition. In the superionic phase, $\sigma$ continues to decrease steadily showing a T$^{-\alpha}$ ($\alpha \approx$~3/2) temperature dependence suggestive of acoustic phonon dominated scattering mechanism in $\beta$-Ag$_2$Te. In samples AGs-300BMS (s = 1, 2, and 4), $\sigma$ near 300 K has reduced to $\rm 1200~S~cm^{-1}$, $\rm 900~S~cm^{-1}$ and $\rm 340~S~cm^{-1}$, respectively. The large reduction for sample AG4-300BMS is particularly striking given that all four samples were prepared under identical conditions and with the same starting stoichiometry. The magnitude of $\sigma$ at 300~K correlates  nicely with the carrier concentration ($n$) at 300~K shown in Table I, which is highest for AG3-300BMS ($\rm 2.4 \times 10^{18}~cm^{-3}$) and lowest for AG4-300BMS ($\rm 0.7 \times 10^{18}~cm^{-3}$). All the samples exhibit high Hall mobilities: $\rm \sim5000~cm^2~V^{-1}~s^{-1}$ for AGs-300BMS (s = 1, 2 and 3) and $\rm \sim3000~cm^2~V^{-1}~s^{-1}$ for AG4-300BMS. These mobility values far exceed those of some better known state-of-the-art n-type thermoelectric materials. For example, the carrier mobility in Bi$_2$Te$_3$ alloys is $\rm \sim200~cm^2~V^{-1}~s^{-1}$ ~\cite{kim2005thermoelectric} or that in Mg$_3$Sb$_2$ alloys is $\rm \sim100~cm^2~V^{-1}~s^{-1}$~\cite{shi2019revelation}. 

The sample dependence is more vividly revealed by the temperature variation of S shown in Fig.~\ref{LSR}b.  While samples AGs-300BMS (s = 1, 2 and 3) show a  n-type behavior, a p-type behavior is observed for AG4-300BMS. This is in line with several previous reports where both n-type and p-type behaviors are found for samples prepared under identical conditions and with same starting molar ratio Ag : Te $\equiv$ 2 : 1. A similar sample dependence has also been seen in the HMS samples. The electrical conductivity and thermopower of two representative HMS samples are also plotted in Fig.~\ref{LSR}(a, b).

The behavior of sintered samples suggests that during the high temperature sintering the sample composition changes due to Ag-ion migration. This indicates that the properties of Ag$_2$Te are highly sensitive to very minor changes in the Ag/Te molar ratio. The high-temperature sintering or spark-plasma sintering (SPS) in the previous works (and the high-T sintering in this work for BMS and HMS samples) is generally done to get high-density pellets. In a recent paper, it was shown that Ag$_2$S exhibit a metal-like ductility in its room-temperature phase with a large plastic deformation strain~\cite{shi2018room}. This metal-like ductility can be exploited to obtain high density pellets directly by cold-pressing. Indeed, the cold-pressed pellets of Ag$_2$Te (as of Ag$_2$S, not shown here) exhibit mass densities lying within $\pm$5\% of the theoretical density. This property is lacking in the Cu$_2$X family, as similar cold-pressed powders of Cu$_2$X resulted in densities in the range $\rm 70\pm 5$\% of the theoretical density as opposed to 100\% for the Ag$_2$X compounds. This eliminates completely the need for high-temperature sintering to enforce densification or SPS to preserve the nanostructuring along with high-density. We further ascertained the highest temperature up to which the Ag diffusion remains local. The FESEM images of two cold-pressed Ag$_2$Te samples that were heated at 575~K (sample 1) and 615~K (sample 2) under a 20~K gradient and a current flow of $\rm \approx 5~A~cm^{-2}$ are shown in Fig.~S6, Supporting Information. While in the sample 2 some Ag aggregation can be  seen at its colder end, sample 1 shows none. This indicates that even within the superionic phase, the Ag migration becomes non-local only at high temperatures, exceeding $\approx$150~K above $\rm T_t$. Based on this, we limited the highest temperature in our experiments to 570~K.

 Equipped with these inputs, we fabricated a series of Ag$_2$Te samples by ball milling followed by cold-pressing (BMC) and measured their TE properties between 300~K and 570~K. Here, we first show our results for BMC samples, AG1-200BMC, AG2-400BMC, AG3-500BMC, AG4-600BMC, and AG5-500BMC. They all have the same starting stoichiometry (molar ratio Ag/Te $\equiv$~2); however, the ball-milling duration has been varied (except between AG3 and AG5) to influence the average  particle-size for reducing the thermal conductivity, discussed later. 

\begin{table*}[!]
	\caption{ Room temperature carrier concentration (n), electrical conductivity ($\sigma$), hall mobility ($\mu$), thermopower (S) and power factor (PF).}
	\label{table 1}
	\centering
	\begin{tabular}{cccccccccc}
		\hline
		\hline
		& & & & & & & &\\
		Samples & n (10$^{18}$ cm$^{-3}$)  & $\sigma$ (Scm$^{-1}$) & $\mu$ (cm$^2$V$^{-1}$s$^{-1}$) & S ($\mu$VK$^{-1}$) & PF ($\mu$Wcm$^{-2}$K$^{-1}$) \\
	
		\hline
		\hline
		AG1-300BMS  & 1.45  & 1233 & 5315 & -103 & 13  \\
		
		AG2-300BMS  & 1  & 868 & 5425 & -103 & 8.9  \\
		
		AG3-300BMS & 2.4 & 1805 & 4700 & -85 & 14.7 \\
		
		AG4-300BMS & 0.7 & 345 & 3080 & -87 & 2.6 \\
		
		AG1-180HMS & 2 & 306 & 956 & -88 & 2.4 \\
		
		AG2-180HMS & 1.2 & 618 & 3219 & -121 & 9.2 \\
		
		AG3-180HMS & 4.8 & 1728 & 2700 & -88 & 12 \\
		
		AG4-180HMS & 3.7 & 1751 & 2958 & -88 & 13.5 \\
		
		AG5-180HMS & 2 & 257 & 3080 & -88 & 1.7 \\
		
		AG7-300BMC-1  & 0.74  & 803 & 2550 & -108 & 3.5  \\
		
		AG6-300BMC-0.5 & 1  & 375 & 2344 & -115 & 5  \\
		
		AG3-500BMC & 3.5 & 2291 & 4091 & -92 & 19.4 \\
		
		AG8-500BMC+0.5 & 3.1 & 2118 & 4270 & -97 & 19.6 \\
		
		AG9-500BMC+1 & 3.2 & 2118 & 4137 & -95 & 19.3 \\
		
		AG1-200BMC  & 3.3  & 2247 & 4256 & -94 & 20 \\
		
		AG2-400BMC  & 3.3 & 2118 & 4013 & -96 & 19.6  \\
		
		AG4-600BMC & 3.5 & 2167 & 3869 & -92 & 18.5 \\
		\hline
	\end{tabular}
\end{table*}
The temperature variation of $\sigma$ and S for these samples is shown in Fig.~\ref{LSR}(d, e). They all exhibit a n-type behavior with their thermopower and electrical conductivity plots overlapping on each other. For comparison data for the Ag$_2$Te sample prepared by SSM is also shown. The milling duration (or the average particle-size) has apparently no measurable effect on the charge transport. These n-type samples have high electrical conductivity ($\rm \sigma \approx 2300~S~cm^{-1}$ near 300~K), which higher than our SSM sample and comparable the best ingot samples  reported previously in literature~\cite{capps2010significant}~ (Fig.~\ref{LSR}(d)). The carrier concentration in all these samples is around $\rm 3.4\pm 0.1 \times 10^{18}~cm^{-3}$, independent of the sample (see Table~I). This shows that the Ag/Te stoichiometry remains intact in these samples leading to a highly reproducible behavior. 

Next, we consider the effect of Ag/Te off-stoichiometry. For this purpose, we synthesized Ag-excess and Ag-deficient samples, namely: AG6-300BMC-0.5, AG7-300BMC-1 (0.5\% and 1\% Ag deficient), AG8-500BMC+0.5, AG9-500BMC+1 (0.5\% and 1\% Ag excess). The temperature variation of $\sigma$ and S for these samples is shown in Fig.~\ref{LSR}(g, h). The data for stoichiometric AG3-500BMC are also included for comparison. The Ag-excess samples show a n-type behavior and their electrical conductivity and thermopower coincides with that for their stoichiometric counterpart AG3-500BMC over the whole temperature range. In stark contrast to this stands the Ag-deficient samples that show a p-type behavior and whose electrical conductivity has been significantly suppressed. For these samples even the qualitative temperature dependence of $\rm \sigma(T)$ has changed both below and above $\rm T_t$. In fact, in the superioinc $\alpha$-phase these samples exhibit an increasing behavior (\rm $d\sigma/dT > 0$), whereas for the n-type samples \rm $d\sigma/dT < 0$. 
The carrier concentration ($n$) in these samples at 300~K is shown in Table~I. In the Ag-excess samples, $\rm n = 3.1\pm0.1 \times 10^{18}$~cm$^{-3}$ is nearly coincident with the carrier cconcentration for the stoichiometric samples. This along with the observation of Ag metal peaks in the Ag-excess samples show that the excess Ag is not incorporated in the lattice. Previously,  Taylor and Wood~\cite{taylor1961thermoelectric} also reported the presence of unreacted silver in their Ag-excess samples prepared by solid-state melting. However, in the Ag-deficient samples, as the Ag vacancy concentration increases, $n$ decreases to $\rm 1.0\pm0.1\times10^{18}~cm^{-3}$ for 0.5\% and $\rm 0.7\pm 0.1\times10^{18}~cm^{-3}$ for the 1\% Ag vacancies. 

The effect of thermal cycling between 300~K and 570~K on $\sigma$ and S has been checked thoroughly. This is shown in in the Fig.~S5, Supporting Information. With the exception of the first heating run in the $\beta$-phase, all the subsequent runs for the 3 measurement cycles overlap each other within the error bars or our measurements. In fact, in the superionic phase, even the data taken during the first heating run overlaps with the cooling data and the data taken in the subsequent runs. The hysteresis in the first run in the $\beta$ phase is not understood as yet, but it could be due to changes in the microstructure upon heating shown in Fig.~S3 as discussed earlier.  

The temperature dependence of PF for our samples is shown in Fig.~\ref{LSR}(c, f, i). As shown in Fig.~\ref{LSR}f, the PF for all our n-type BMC samples ( $\rm \approx 20~\mu W~cm^{-1}K^{-2}$ at 300 K) overlap over the whole temperature range. In contrast, in the sintered BMS or HMS samples (Fig.~\ref{LSR}c), the PF is not only low ( $\rm < 12~\mu W~cm^{-1}K^{-2}$ at 300 K) but the PF for these samples is also varies from one sample to other. This underpins the strength of the sample fabrication technique reported here. In our n-type BMC samples, the maximum PF obtained is close to $\rm 20~\mu W~cm^{-1}K^{-2}$ in the $\beta$-phase and $\rm 11~\mu W~cm^{-1}K^{-2}$ in the superionic phase where it shows a weak temperature dependence. In the p-type BMC samples the PF shows a linearly increasing behavior in the superionic phase attaining a value of $\rm 6~\mu W~cm^{-1}K^{-2}$ at 570~K (Fig.~\ref{LSR}i).

To gain a qualitative understanding of the charge transport properties in Ag$_2$Te samples, we turn now to the electronic band structure calculations. The band structure for $\beta$-Ag$_2$Te is shown in Fig.~S6 where an indirect bandgap of 0.07~eV has been observed upon incorporating the spin-orbit coupling in good agreement with previous reports~\cite{zhang2011topological,jahangirli2018ab}. The small band gap suggests that both types of charge carriers contribute to the transport. However, since the carrier mobility ratio $\rm \mu_e/ \mu_h$ in the $\beta$-phase is about 6~\cite{taylor1961thermoelectric}, the sign of S is negative. The band structure for $\beta$-Ag$_2$Te with 6.5\% Ag vacancies is shown in Fig.~S6. Due to Ag vacancies the Fermi level has been pushed down into the valence band. This should have resulted in a positive Seebeck coefficient; however, the amount of Ag vacancies in our samples is less than or equal to 1\%, which is not large enough to make S positive. This is depicted with the help of Fermi surface plots corresponding to the experimental carrier concentrations. The electronic band structure for $\alpha$-Ag$_2$Te is shown in Fig.~S7, which agrees favorably with previous literature~\cite{zhang2011topological}. The Fermi surface in this case consists of a small hole-pocket centered at the $\Gamma$-point and six equivalent electron pockets around the $\rm L$-point. This is expected to give rise to a n-type behavior for stoichiometric $\alpha$-Ag$_2$Te as is indeed seen experimentally. In the 6.5\% Ag-deficient sample, the Fermi level goes deeper into the valence band resulting in a large hole-Fermi surface centered around the $\Gamma$-point as shown in Fig.~S7 which is expected to result in a p-type behavior as is indeed seen experimentally in the $\alpha$-phase of Ag-deficient samples. 

\subsection{Thermal transport}
\begin{figure*}[!]
	\includegraphics[width=17.5cm]{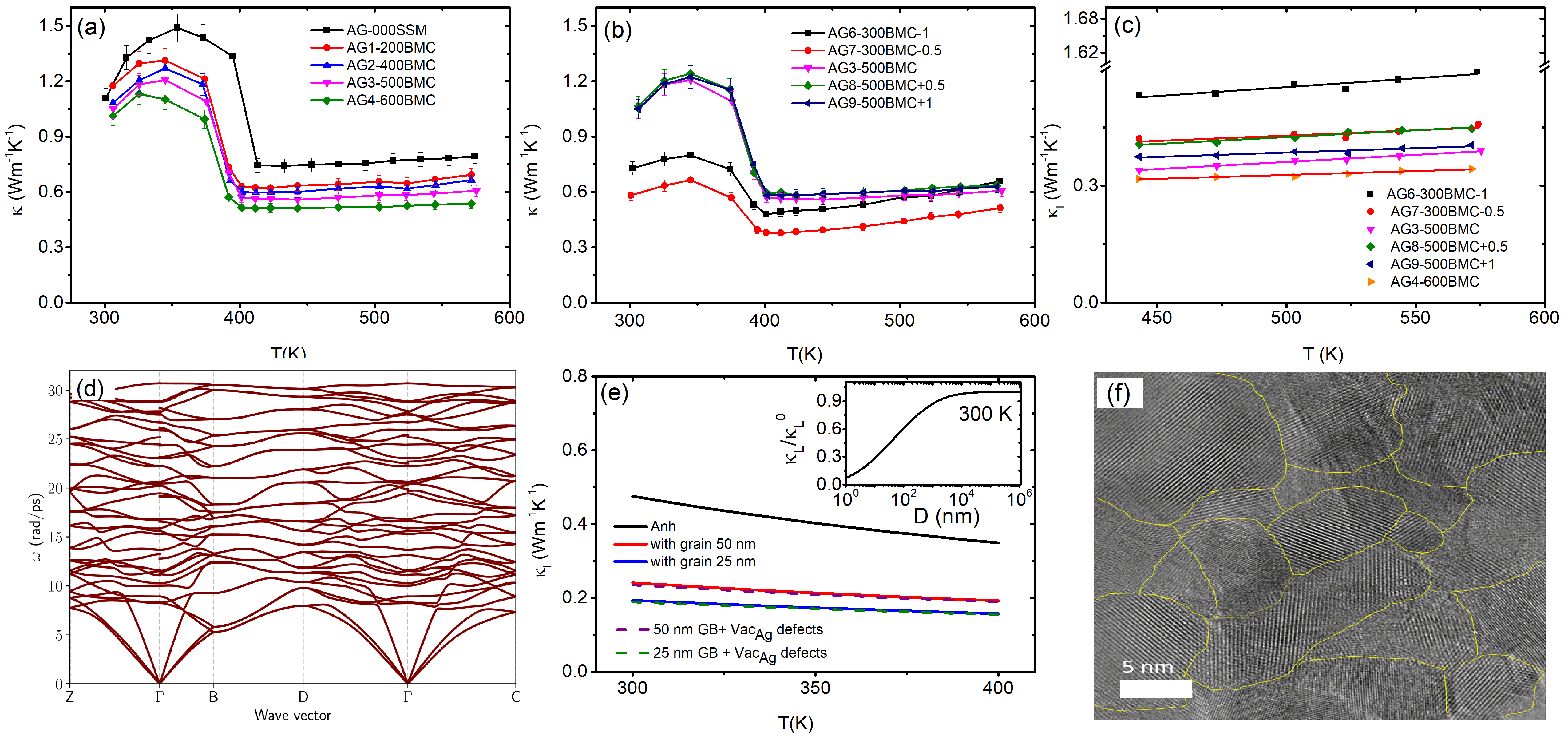}
	\caption{(a) Temperature dependent $\kappa$ of BMC samples prepared with different milling durations, (b) Temperature dependent $\kappa$ of BMC samples prepared with different Ag/Te molar ratio, (c) $\kappa_l$ extracted from the measured $\kappa$ using the SKB mode for a few samples, (d) Calculate phonon dispersion for $\beta$-Ag$_2$Te, (d) Theoretically calculated thermal conductivity showing effect of anharmonicity, grain-size and Ag-vacancies/Te-interstitials, (e) a representative high-resolution TEM image of a BMC Ag$_2$Te sample after several thermal cycling. In (a), thermal conductivity of our ingot sample is also included for comparison.}
	\label{k and zT}
\end{figure*}

The thermal conductivity ($\kappa$) of our BMC samples is shown in Fig.~\ref{k and zT}(a, b). The data for SSM sample is also included as a reference. $\kappa$ for remaining samples is shown in Fig.~S8, Supporting Information. 
In the context of Cu$_2$S, it was previously reported that estimating $\kappa_e$ in these intrinsically low $\kappa$ materials often yield unsatisfactory results~\cite{ge2016high}. We estimated $\kappa_e$ using the SKB model as done previously for PbTe-doped n-type Ag$_2$Te samples. The details are shown in the Supporting information. However, the exact determination of $\kappa_e$ is not eminently necessary when comparing $\kappa$ in our stoichiomteric BMC samples as their electrical conductivities overlaps over the whole temperature range. Hence, the $\kappa_e$ contribution in these samples is expected to be the same. Thus, a comparison of the as measured $\kappa$ for these samples with variable milling time is sufficient to understand the effect on the lattice thermal conductivity.

In all the n-type samples, $\kappa$ shows the same qualitative behavior characterized by a broad peak centered around 340~K, followed by a precipitous drop across $\rm T_t$ and a gradually increasing trend ($\rm d\kappa/dT > 0$) upon heating up to the highest temperature in the superionic phase. The fact that $\kappa$ in our SSM sample is significantly higher compared to $\kappa$ for any of the BMC samples despite having a much lower electrical conductivity ($\rm 1500~S~cm^{-1}$ near 300~K in SSM against $\rm 2300~S~cm^{-1}$ near 300~K in BMC, Fig.~\ref{LSR}c) clearly brings out the role of grain boundary scattering in suppressing the lattice thermal conductivity. With increasing milling duration, $\kappa$ decreases further in a systematic manner showing the effect of nanostructuring in suppressing the lattice thermal conductivity of Ag$_2$Te drastically. The lowest $\kappa$ is observed for the sample that was milled for the longest duration (600 min). For this sample sample (AG4-600BMC, Fig.~\ref{k and zT}a), the maximum and minimum values of $\kappa$ are, respectively, $\rm 1.1~W~m^{-1}~K^{-1}$ (at 350~K) and $\rm 0.47~W~m^{-1}~K^{-1}$ (at 420~K). The corresponding values for our SSM sample are $\rm 1.5~W~m^{-1}~K^{-1}$ and $\rm 0.75~W~m^{-1}~K^{-1}$, respectively. The broad peak in $\kappa$ above room temperature in the $\beta$-phase is reminiscent of the peak at the same temperature in $\sigma$. On the other hand, the gradual increase in the superionic phase ($\rm d\kappa/dT > 0$) is contrary to the decreasing behavior shown by $\sigma$ ($\rm d\sigma/dT < 0$) over the same temperature range. This anomalous temperature dependence of $\kappa$ has recently attracted significant attention. It is believed that the heat transport in PLEC class of materials cannot be described within the classical paradigm of lattice vibrations alone. Wu et al., argued that the anomalous behavior is associated with the convective heat channel due to the mobile Ag ions, in addition to the conductive heat channel due to lattice vibrations of Te atoms~\cite{wu2018two}. 

The effect of Ag off-stoichiometry is shown in Fig.~\ref{k and zT}b. $\kappa$ for pristine, 0.5\% and 1\% Ag-excess samples overlap. This observation is consistent with the conclusion that excess Ag is not incorporated in the structure, drawn first on the basis of PXRD (extra peak due to unreacted Ag metal) and then the charge transport data (unchanged $\sigma$ and S). Compared to these samples, the thermal conductivity of 0.5\% and 1\% Ag deficient samples has decreased mainly due to their very low electrical conductivity (Fig.~\ref{LSR}g).  In Fig.~\ref{k and zT}c, we show $\kappa_l$ for these samples in the superionic phase (see, Supporting Information for details). Data for AG4-600BMC samples is also included for comparison. $\kappa_l$, as expected, is lowest for AG4-600BMC ($\rm \approx 0.35~W~m^{-1}~K^{-1}$), but unexpectedly the highest lattice thermal conductivity is shown by 1\% Ag-deficient sample. This shows that the atomic defects are not the main characters in the heat flow mechanism of Ag$_2$Te. For all samples $\rm d\kappa_l/dT > 0$, which, as we already pointed out above, is believe to be due to the convective heat transport due to mobile Ag-ions.

Interestingly, the thermal conductivity of Ag$_2$Te is also very low in the low-tempertaure $\beta$-phase. To understand the origin of this intrinsically low thermal conductivity in the $\beta$-phase of Ag$_2$Te, we studied the effect of anharmonicity, grain boundary and point-defects scattering theoretically. The phonon dispersion in $\beta$- Ag$_2$Te is shown in  Fig.~\ref{k and zT}d (see also Figure~S11, Supporting Information). There are 36 phonon branches at each q-point, spanning a small energy range from 0 to 30 rad/ps. This is due to the large average molecular mass of Ag$_2$Te (114~amu). The low lying optical phonons are prominently seen without an acoustic-optic phonon gap due to nearly similar atomic masses of Ag and Te ($\rm m_{Ag}/m_{Te} = 0.85)$. Both these factors contribute to a high intrinsic anharmonic phonon-phonon scattering in Ag$_2$Te. Furthermore, less dispersive phonon branches indicate low phonon group velocities. Thus, a combined effect of these is expected to result in an inherently low $\kappa_l$ in this material. 

To get further insight, different phonon scattering contributions in $\beta$-Ag$_2$Te are calculated. The details are given in the  Supporting Information. The anharmonic scattering rate ($\rm \tau^{-1}_{anh}$) is calculated using the method in Ref.~\cite{madsen2016calculating}. 
As anticipated from the phonon dispersion, the anharmonic scattering in Ag$_2$Te is high, which is evident in Fig.~S12b. The grain boundary scattering is included as $\rm \tau^{-1}_{GB} = v_g/D$ where $\rm v_g$ is the phonon group velocity and D is the grain size. The effects of Ag vacancy ($\rm Vac_{Ag}$) is included using the T-matrix description as discussed in Ref.~\cite{katre2017exceptionally}. The defect scattering rates, $\tau^{-1}_{\text{def}}$, are seemingly the least dominant among all the contributions and could be anticipated to have less impact on calculated $\kappa_l$. A full ab-initio treatment of phonon-defect interactions, taking into account the bond perturbations also, may somewhat strengthen $\rm \tau^{-1}_{def}$, however, its final effect on $\kappa_l$ may still remain less pronounced due to already very high anharmonic and grain boundary scattering, see Fig.~S12b. The total scattering rate ($\rm \tau^{-1}$) is calculated from full phonon dispersion and mode Gr\"uneisen parameters using the Matheissen rule: $\rm \tau^{-1} = \tau^{-1}_{anh} + \tau^{-1}_{GB} + \tau^{-1}_{def}$. Fig.~\ref{k and zT}c shows the calculated $\kappa_l$ for Ag$_2$Te including all the phonon scattering contributions. The calculated $\kappa_l$ for Ag$_2$Te (Fig.~\ref{k and zT}e), considering only  the anharmonicity to begin with, is already very low ($\rm \sim 0.45~W~m^{-1}~K^{-1}$ at 300~K). Including the grain boundary scattering, taking grain-size $L =$ 50~nm and 25~nm, reduces $\kappa_l$ further by more than 50\%, which is consistent with the observed low thermal conductivity of our nanostructured Ag$_2$Te samples.  The effect of small concentrations of Ag vacancies on $\kappa$ is relatively very weak. This is consistent with the experiments where the Ag-deficiency has not contributed to any further decrease in thermal conductivity.

A more detailed view of the normalised $\kappa_l$ at room-temperature varying with the grain size is shown in the inset of Fig.~\ref{k and zT}e. This shows that grains, sizing below 1~$\mu$m, contribute to the suppression of $\kappa$ and the smaller grains are most effective in scattering the mid- and long-wavelength phonons. As elucidated in Fig.~S4c and Fig.~S3, our samples consists of grains that vary in size from 1~$\mu$m down to the less than 100~nm. These grains, however, are themselves composed of nm sized particles of Ag$_2$Te shown in Fig.~\ref{TEM} which cannot be seen in the SEM micrograph. We therefore did extensive HRTEM investigations on our BMC samples before and after the thermoelectric measurements. A representative HRTEM image collected after the thermoelectric measurements is shown in Fig.~\ref{k and zT}f (a few more TEM images are shown in the Supporting Information, Fig.~S5). One can clearly see the presence of a large number of nm-sized grains oriented randomly. The grain boundaries are drawn as lines as a guide to eye. These boundaries are not very sharp everywhere, neither do all grains show high-crystallinity, acting thereof as potential scatterers that impede the heat flow. It is well recognized that a wide size distribution and morphology of nanoparticles is more effective in scattering different phonon modes and reduce thermal conductivity significantly. The presence of grains of varying size with a broad distribution in our cold-pressed pellets thus scatter phonons at all length scales leading to their much reduced thermal conductivity compared to an ingot sample.     
    
\subsection{Thermoelectric figure-of-merit zT}
\begin{figure*}[!]
	\includegraphics[width=16cm]{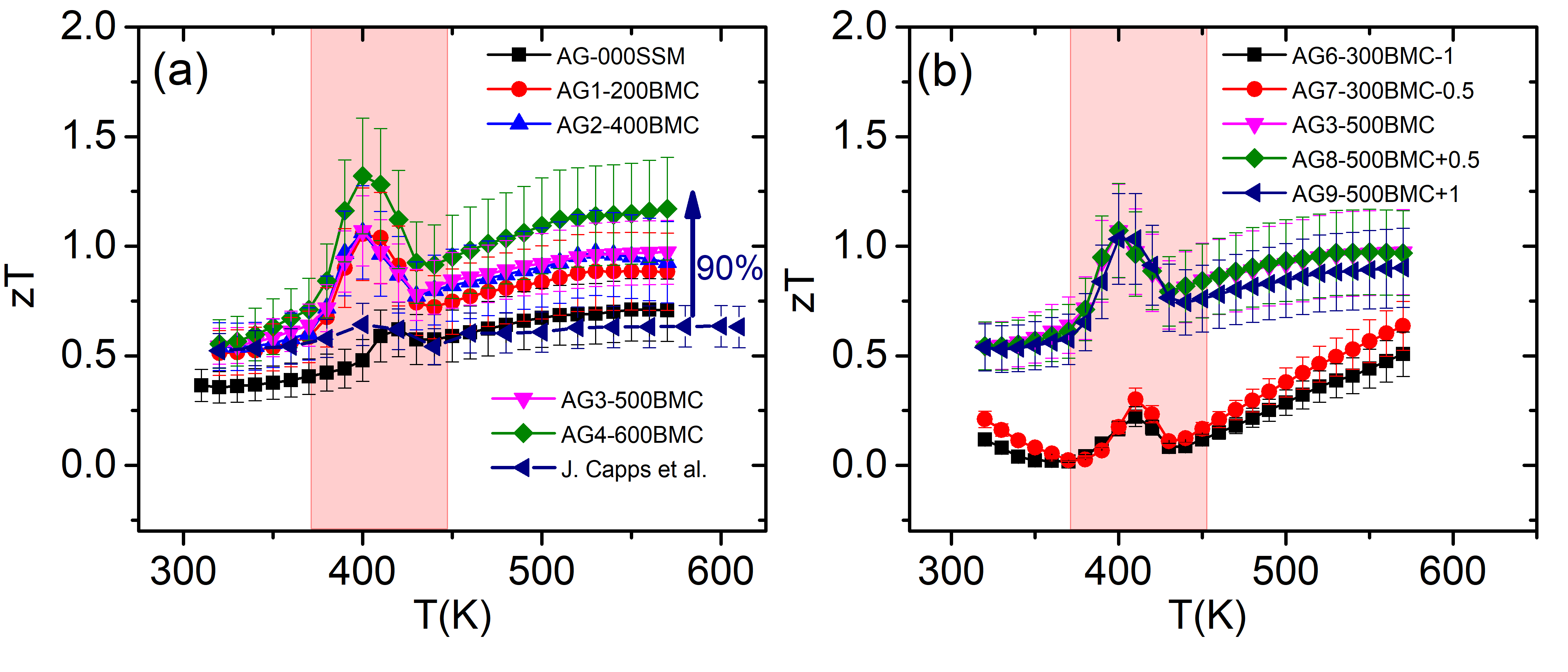}
	\caption{(Temperature dependent zT variation for (a) stoichiomteric BMC samples prepared with different milling times (SSM sample and literature data from J. Capps et al. study are included for comparison), and (b) different Ag/Te molar ratio receptively.}
	\label{zT}
\end{figure*}

The zT of our BMC samples is plotted as a function of temperature in Fig.~\ref{zT}(a, b). In panel (a), the zT for SSM sample is also included as a reference. For the BMS and HMS samples, zT is shown in Fig.~S8b, Supporting Information. Since the specific heat near the phase transition at $\rm T_t$ is expected to show an anomaly, we exclude the zT in the range $\rm T_t \pm 100~K$ (the shaded region) from further discussion. We first consider the n-type samples where zT shows an increasing trend over the whole temperature range. In BMS or HMS samples, the highest zT obtained is 0.85 at 570~K. In contrast to this, in the n-type BMC samples the highest zT of 1.2 is obtained for the sample with the lowest thermal conductivity.

 In the BMC samples, the zT shows a systematic increases with decrease in the thermal conductivity due to increasing milling time. Since ball-milling affects the particle-size and their morphologies, the observed suppression reflects the effect of nanostructuring in these materials. Interestingly, the ball-milling has no significant effect on the electrical conductivity and thermopower. This makes Ag$_2$Te a rather interesting material where the power factor and phonon transport are intrinsically decoupled, providing thereby a huge scope for further improvement.  In contrast the zT of SSM sample is only 0.68 at 570~K. This low value is mainly due to its fairly high thermal conductivity compared to the BMC samples. The effect of Ag/Te molar ratio on zT of the BMC samples is shown in Fig.~\ref{zT}b. In the Ag-excess samples, the zT values nearly coincides with their stoichiometric counterpart. In the Ag-deficient, p-type BMC samples, zT shows a decreasing behavior below room-temperature but an increasing behavior in the superionic phase with zT reaching values as high as 0.64 at 570~K~(Fig.~\ref{zT}b). On the other hand, in the p-type sintered samples zT is very small and it also shows a decreasing behavior in the superionic phase (Fig.~S8b).

\section{SUMMARY AND CONCLUSIONS}
The thermoelectric properties of Ag$_2$Te samples, synthesized using four different methods, namely, BMS, HMS, BMC and SSM were investigated in detail. The sintered BMS or HMS samples prepared under identical conditions and with the same starting stoichiometry (Ag : Te $\equiv$ 2 : 1) show highly unpredictable behavior with respect to the sign of S and the magnitude of $\sigma$. The leading reason for this extreme sample dependence is related to the Ag-ion diffusion during high-temperature sintering. To over come this issue and to maximize the zT of pristine Ag$_2$Te, we developed a novel, all-room-temperature, fabrication technique that involves vibratory ball milling followed by cold-pressing. The cold pressed pellets were highly dense (within $\pm5$\% of the theoretical density). We demonstrate excellent reproducibility of the thermoeelctric properties from one sample to another. To improve the zT of Ag$_2$Te beyond the phonon-liquid electron-crystal value of about 0.64 at 575~Kin the ingot samples, we adopted a heirarchical nanostructuring technique, wherein nanoparticles with diverse morphology and crystallinity (nanoscale), along wih larger particles (mesoscale) were used. This effectively suppressed the thermal conductivity of our samples drastically leading to a drastic improvement in the zT values for both n-type and p-type Ag$_2$Te: 1.2 in n-type and 0.64 in p-type at 570~K. This temperature is much than the temperature where the Cu-based superionic materials attain their peak zT. In future, charge carrier optimization by doping in Ag$_2$Te using the fabrication technique described here can be considered to achieve higher zT, and to suppress Ag-migration up to even higher temperatures than 570 K shown here for the pristine Ag$_2$Te. As our fabrication technique for obtaining 100\% dense pellets does not involve high-temperature sintering or spark-plasma sintering, it can readily be scaled-up for higher yields.

\section{METHODS}
\label{ExpCalc}
\subsection{Experimental details}
\label{Exp} 
Polycrystalline samples of Ag$_2$Te were synthesized by four different methods, namely BMS (Ball-milled and sintered), HMS (Hand-milled and sintering), BMC (Band milled and cold pressed) and SSM (solid state melting). In all cases, Ag powders from Sigma Aldrich (purity 99.98\%) and Te powder from Alfa Aesar (purity 99.999\%) were used. The purity and source of precursors was maintained throughout to avoid any variations among the synthesized samples due to unknown factors. For the synthesis of ball-milled Ag$_2$Te nanopowders, the elemental powders were weighed in the molar ratio according to the desired stoichiometry (Ag/Te $\equiv$ 2 for stoichiometric and $\rm 2 \pm \delta$ for the off-stoichiometric samples) and transferred into the stainless steel (SS) cylindrical vessel of the milling apparatus along with the SS balls. The vessel was than closed with a lid having a teflon gasket to preserve sample atmosphere during ball-milling. This steps was carried out in an Ar-filled glove-box (Mbraun) to avoid any oxidation. The loaded vessel was then removed from the glove-box and mounted on the vibratory ball mill. The volume of the vessel used was 6~cm$^3$. The SS beads used were of diameter 0.4~cm. The sample to balls mass ratio was kept around 1 : 2.5. The vessel oscillates along the arc of circle. The vibration frequency was 120 rpm, and amplitude of vibration around $\rm \pm 10^\circ$ about the mean vertical position. The ball-milled powders were cold pressed in a in KBr press die set to get the BMC series of samples. For the BMS series, the cold-pressed pellets were flame sealed in quartz ampoules under dynamic vacuum and sintered at 673~K for 24 hours in a muffle furnace with heating elements on the two sides (which is the potential source of temperature gradient leading to Ag migration). The HMS samples were prepared by hand-grinding the Ag and Te powders for 180 minutes inside an Ar-filled glove box. The samples were subsequently cold-pressed in a KBr die-set and sintered in evacuated quartz ampoules as describes above. The SSM sample was prepared directly by melting Ag and Te powder in an evacuated quartz ampule. The ampoule was loaded in a muffle furnace and heated to 1000$\rm ^\circ$C in 10 h, kept at this temperature for 5 hours and then cooled down to 500$\rm ^\circ$C at a rate of 10$\rm ^\circ$C h$^{-1}$. The furnace was then turned-off to allow cooling down to the room temperature. The phase purity of our samples was assessed using the x-ray powder diffraction technique (Bruker, D8 Advance). The lattice parameters were estimated by performing the Rietveld refinement using the Fullprof Suite. The microstructure was assessed using the Field Effect Scanning Electron Microscopy (FESEM) (Ultra Zeiss plus) equipped with an Energy Dispersive X-ray (EDX) analysis tool (Oxford Instruments). The high-resolution transmission electron microscopy (HRTEM) (JEOL JEM 2200FS 200 KV) was done to examine the microstructure. For this purpose, the ball-milled or finely ground powders were mixed with ethanol in a vial which was then centrifuged and the smaller particles from the top of the vial were dropcasted onto a Cu-grid that was subsequently loaded on the TEM sample stage and plasma cleaned at 250~eV. The GSM-3 software package was used for further FFT and IFFT analysis. The electrical resistivity and thermopower were measured simultaneously using Linseis LSR-3 set-up. The thermal diffusivity was measured on disk shaped samples ($\rm \phi = 8$~mm) using the Laser Flash Analyzer (Linseis LFA 100). The thermal conductivity is then obtained by using the formula, $\rm \kappa = C_p \rho_mD$ where $\rm C_p$ is the specific heat at constant pressure, $\rho_m$ is sample density and D is the measured thermal diffusivity. The C$_p$ was approximated by its Dulong-Petit limit (C$_p$ = 3nR), which is incorrect around the phase transition where the specific heat shows an anomaly. For this reason, the temperature range  $\rm T_t \pm~100 K$ has been excluded from the analysis of zT. Sample density was estimated by measuring the mass and dimensions of disk shaped samples. The measured values were cross-checked using the Archimedes method for several samples and the two methods agreed within $\rm \pm 5\%$. Room temperature (RT) Hall carrier concentration was measured using a Physical Property Measurement System (Quantum Design). 

\subsection{Computational Details}
\label{Comp}
The first principles DFT calculations were performed using the Quantum Espresso code (QE)~\cite{giannozzi2009quantum}. The exchange correlation energy is represented by a generalized gradient approximation (GGA) functional proposed by Perdew, Burke and Ernzerhof (PBE)~\cite{perdew1996generalized}. The mess cut-off for the plane wave was kept as 50 Ry for the ground state calculations. The crystal structure of the room-temperature $\beta$-Ag$_2$Te is monoclinic with 4 fu per unit cell. The Brillouin zone is sampled using a Monkhorst-Pack scheme, a $\rm 12 \times24 \times 12$ and $\rm 20 \times 20 \times 20$ mesh of k-point is used for the geometry optimization. Density of states (DOS) and band structure calculations were performed using a more denser grid of $\rm 12 \times24 \times 12$ and $\rm 26 \times26 \times 26$. The structure is relaxed until the force and energy convergence criteria of 10$^{-6}$~eV and 10$^{-7}$ eV~\AA$^{-1}$, respectively, are achieved. The lattice parameter for the $\beta$ phase are, $\rm~a = 8.30~\text{\r{A}}$, $\rm~b = 4.47~\text{\r{A}}$, $\rm~c = 7.53~\text{\r{A}}$ and $\rm~a = 6.70~\text{\r{A}}$ for superionic cubic $\alpha$-phase. These geometrical parameters are in reasonable agreement with the theoretically reported values \cite{zhang2011topological}. The band structures in both the phases were also calculated for 6.5\% Ag vacancy. For this, we considered a $\rm 1 \times 2 \times 1$ and $\rm 2 \times 2 \times 2$ supercell of Ag$_2$Te for the $\beta$ and $\alpha$ phase, respectively. The DOS calculations for supercell are performed using a k-mesh of $\rm 12 \times 12 \times 12$ and $\rm 10 \times 10 \times 10$ for the $\beta$ and $\alpha$ phase. The thermal transport properties of $\beta$-Ag$_2$Te are calculated within ab initio framework from complete phonon description. The normal modes of phonons are calculated using finite displacement approach for $\rm 2 \times 4 \times 2$ supercell to avoid periodic boundary effects on calculated atomic forces. LO-TO splitting is accounted by including non-analytical term correction, for which the dielectric tensor and Born effective charges (BEC) are required. These properties along with the atomic forces for each supercell with displacement are calculated using QE package, whereas the dynamical matrix extraction and diagonalization is performed using Phonopy~\cite{togo2015first}. Anharmonicity is considered within Quasi harmonic approximation (QHA) that required phonon calculations on $\pm$3 \% of relaxed cell volume. GGA(PBE) and LDA (PZ) \cite{perdew1996generalized} exchange-correlation functions, both are tested for computing phonon frequencies and LDA results are further used to determine different phonon scattering rate ($\tau$) contributions and the lattice thermal conductivity $\kappa$$_l$.

\medskip
\textbf{Acknowledgements} \par 
NJ is thankful to CSIR for the PhD fellowship. SS acknowledge financial support from SERB under research grant no. CGR/2021/005804. AK acknowledges support from DST-INSPIRE Faculty Scheme (Grant No. IFA17-MS122).  
\medskip

\medskip
\textbf{Authors contribution} \par 
The sample preparation, their structural characterization and thermoelectric properties measurements were carried out by NJ under the guidance of SS. The theoretical calculations were done by NB and AK under the guidance of AK. The project was conceived and executed by SS. The manuscript was written by NJ and SS with inputs from all the authors.

\bibliography{mybib.bib}

\begin{thebibliography}{60}%
\makeatletter
\providecommand \@ifxundefined [1]{%
 \@ifx{#1\undefined}
}%
\providecommand \@ifnum [1]{%
 \ifnum #1\expandafter \@firstoftwo
 \else \expandafter \@secondoftwo
 \fi
}%
\providecommand \@ifx [1]{%
 \ifx #1\expandafter \@firstoftwo
 \else \expandafter \@secondoftwo
 \fi
}%
\providecommand \natexlab [1]{#1}%
\providecommand \enquote  [1]{``#1''}%
\providecommand \bibnamefont  [1]{#1}%
\providecommand \bibfnamefont [1]{#1}%
\providecommand \citenamefont [1]{#1}%
\providecommand \href@noop [0]{\@secondoftwo}%
\providecommand \href [0]{\begingroup \@sanitize@url \@href}%
\providecommand \@href[1]{\@@startlink{#1}\@@href}%
\providecommand \@@href[1]{\endgroup#1\@@endlink}%
\providecommand \@sanitize@url [0]{\catcode `\\12\catcode `\$12\catcode
  `\&12\catcode `\#12\catcode `\^12\catcode `\_12\catcode `\%12\relax}%
\providecommand \@@startlink[1]{}%
\providecommand \@@endlink[0]{}%
\providecommand \url  [0]{\begingroup\@sanitize@url \@url }%
\providecommand \@url [1]{\endgroup\@href {#1}{\urlprefix }}%
\providecommand \urlprefix  [0]{URL }%
\providecommand \Eprint [0]{\href }%
\providecommand \doibase [0]{https://doi.org/}%
\providecommand \selectlanguage [0]{\@gobble}%
\providecommand \bibinfo  [0]{\@secondoftwo}%
\providecommand \bibfield  [0]{\@secondoftwo}%
\providecommand \translation [1]{[#1]}%
\providecommand \BibitemOpen [0]{}%
\providecommand \bibitemStop [0]{}%
\providecommand \bibitemNoStop [0]{.\EOS\space}%
\providecommand \EOS [0]{\spacefactor3000\relax}%
\providecommand \BibitemShut  [1]{\csname bibitem#1\endcsname}%
\let\auto@bib@innerbib\@empty
\bibitem [{\citenamefont {Bell}(2008)}]{bell2008cooling}%
  \BibitemOpen
  \bibfield  {author} {\bibinfo {author} {\bibfnamefont {L.~E.}\ \bibnamefont
  {Bell}},\ }\bibfield  {title} {\bibinfo {title} {Cooling, heating, generating
  power, and recovering waste heat with thermoelectric systems},\ }\href@noop
  {} {\bibfield  {journal} {\bibinfo  {journal} {Science}\ }\textbf {\bibinfo
  {volume} {321}},\ \bibinfo {pages} {1457} (\bibinfo {year}
  {2008})}\BibitemShut {NoStop}%
\bibitem [{\citenamefont {He}\ and\ \citenamefont
  {Tritt}(2017)}]{he2017advances}%
  \BibitemOpen
  \bibfield  {author} {\bibinfo {author} {\bibfnamefont {J.}~\bibnamefont
  {He}}\ and\ \bibinfo {author} {\bibfnamefont {T.~M.}\ \bibnamefont {Tritt}},\
  }\bibfield  {title} {\bibinfo {title} {Advances in thermoelectric materials
  research: Looking back and moving forward},\ }\href@noop {} {\bibfield
  {journal} {\bibinfo  {journal} {Science}\ }\textbf {\bibinfo {volume}
  {357}},\ \bibinfo {pages} {eaak9997} (\bibinfo {year} {2017})}\BibitemShut
  {NoStop}%
\bibitem [{\citenamefont {Sootsman}\ \emph {et~al.}(2009)\citenamefont
  {Sootsman}, \citenamefont {Chung},\ and\ \citenamefont
  {Kanatzidis}}]{sootsman2009new}%
  \BibitemOpen
  \bibfield  {author} {\bibinfo {author} {\bibfnamefont {J.~R.}\ \bibnamefont
  {Sootsman}}, \bibinfo {author} {\bibfnamefont {D.~Y.}\ \bibnamefont
  {Chung}},\ and\ \bibinfo {author} {\bibfnamefont {M.~G.}\ \bibnamefont
  {Kanatzidis}},\ }\bibfield  {title} {\bibinfo {title} {New and old concepts
  in thermoelectric materials},\ }\href@noop {} {\bibfield  {journal} {\bibinfo
   {journal} {Angewandte Chemie International Edition}\ }\textbf {\bibinfo
  {volume} {48}},\ \bibinfo {pages} {8616} (\bibinfo {year}
  {2009})}\BibitemShut {NoStop}%
\bibitem [{\citenamefont {Zide}\ \emph {et~al.}(2006)\citenamefont {Zide},
  \citenamefont {Vashaee}, \citenamefont {Bian}, \citenamefont {Zeng},
  \citenamefont {Bowers}, \citenamefont {Shakouri},\ and\ \citenamefont
  {Gossard}}]{zide2006demonstration}%
  \BibitemOpen
  \bibfield  {author} {\bibinfo {author} {\bibfnamefont {J.}~\bibnamefont
  {Zide}}, \bibinfo {author} {\bibfnamefont {D.}~\bibnamefont {Vashaee}},
  \bibinfo {author} {\bibfnamefont {Z.}~\bibnamefont {Bian}}, \bibinfo {author}
  {\bibfnamefont {G.}~\bibnamefont {Zeng}}, \bibinfo {author} {\bibfnamefont
  {J.}~\bibnamefont {Bowers}}, \bibinfo {author} {\bibfnamefont
  {A.}~\bibnamefont {Shakouri}},\ and\ \bibinfo {author} {\bibfnamefont
  {A.}~\bibnamefont {Gossard}},\ }\bibfield  {title} {\bibinfo {title}
  {Demonstration of electron filtering to increase the seebeck coefficient in
  in 0.53 ga 0.47 as/ in 0.53 ga 0.28 al 0.19 as superlattices},\ }\href@noop
  {} {\bibfield  {journal} {\bibinfo  {journal} {Physical Review B}\ }\textbf
  {\bibinfo {volume} {74}},\ \bibinfo {pages} {205335} (\bibinfo {year}
  {2006})}\BibitemShut {NoStop}%
\bibitem [{\citenamefont {Faleev}\ and\ \citenamefont
  {L{\'e}onard}(2008)}]{faleev2008theory}%
  \BibitemOpen
  \bibfield  {author} {\bibinfo {author} {\bibfnamefont {S.~V.}\ \bibnamefont
  {Faleev}}\ and\ \bibinfo {author} {\bibfnamefont {F.}~\bibnamefont
  {L{\'e}onard}},\ }\bibfield  {title} {\bibinfo {title} {Theory of enhancement
  of thermoelectric properties of materials with nanoinclusions},\ }\href@noop
  {} {\bibfield  {journal} {\bibinfo  {journal} {Physical Review B}\ }\textbf
  {\bibinfo {volume} {77}},\ \bibinfo {pages} {214304} (\bibinfo {year}
  {2008})}\BibitemShut {NoStop}%
\bibitem [{\citenamefont {Heremans}\ \emph {et~al.}(2008)\citenamefont
  {Heremans}, \citenamefont {Jovovic}, \citenamefont {Toberer}, \citenamefont
  {Saramat}, \citenamefont {Kurosaki}, \citenamefont {Charoenphakdee},
  \citenamefont {Yamanaka},\ and\ \citenamefont
  {Snyder}}]{heremans2008enhancement}%
  \BibitemOpen
  \bibfield  {author} {\bibinfo {author} {\bibfnamefont {J.~P.}\ \bibnamefont
  {Heremans}}, \bibinfo {author} {\bibfnamefont {V.}~\bibnamefont {Jovovic}},
  \bibinfo {author} {\bibfnamefont {E.~S.}\ \bibnamefont {Toberer}}, \bibinfo
  {author} {\bibfnamefont {A.}~\bibnamefont {Saramat}}, \bibinfo {author}
  {\bibfnamefont {K.}~\bibnamefont {Kurosaki}}, \bibinfo {author}
  {\bibfnamefont {A.}~\bibnamefont {Charoenphakdee}}, \bibinfo {author}
  {\bibfnamefont {S.}~\bibnamefont {Yamanaka}},\ and\ \bibinfo {author}
  {\bibfnamefont {G.~J.}\ \bibnamefont {Snyder}},\ }\bibfield  {title}
  {\bibinfo {title} {Enhancement of thermoelectric efficiency in pbte by
  distortion of the electronic density of states},\ }\href@noop {} {\bibfield
  {journal} {\bibinfo  {journal} {Science}\ }\textbf {\bibinfo {volume}
  {321}},\ \bibinfo {pages} {554} (\bibinfo {year} {2008})}\BibitemShut
  {NoStop}%
\bibitem [{\citenamefont {Wu}\ \emph {et~al.}(2017)\citenamefont {Wu},
  \citenamefont {Li}, \citenamefont {Wang}, \citenamefont {Zhang},
  \citenamefont {Yang}, \citenamefont {Zhang}, \citenamefont {Chen},\ and\
  \citenamefont {Yang}}]{wu2017resonant}%
  \BibitemOpen
  \bibfield  {author} {\bibinfo {author} {\bibfnamefont {L.}~\bibnamefont
  {Wu}}, \bibinfo {author} {\bibfnamefont {X.}~\bibnamefont {Li}}, \bibinfo
  {author} {\bibfnamefont {S.}~\bibnamefont {Wang}}, \bibinfo {author}
  {\bibfnamefont {T.}~\bibnamefont {Zhang}}, \bibinfo {author} {\bibfnamefont
  {J.}~\bibnamefont {Yang}}, \bibinfo {author} {\bibfnamefont {W.}~\bibnamefont
  {Zhang}}, \bibinfo {author} {\bibfnamefont {L.}~\bibnamefont {Chen}},\ and\
  \bibinfo {author} {\bibfnamefont {J.}~\bibnamefont {Yang}},\ }\bibfield
  {title} {\bibinfo {title} {Resonant level-induced high thermoelectric
  response in indium-doped gete},\ }\href@noop {} {\bibfield  {journal}
  {\bibinfo  {journal} {NPG Asia Materials}\ }\textbf {\bibinfo {volume} {9}},\
  \bibinfo {pages} {e343} (\bibinfo {year} {2017})}\BibitemShut {NoStop}%
\bibitem [{\citenamefont {Liu}\ \emph {et~al.}(2012)\citenamefont {Liu},
  \citenamefont {Tan}, \citenamefont {Yin}, \citenamefont {Liu}, \citenamefont
  {Tang}, \citenamefont {Shi}, \citenamefont {Zhang},\ and\ \citenamefont
  {Uher}}]{liu2012convergence}%
  \BibitemOpen
  \bibfield  {author} {\bibinfo {author} {\bibfnamefont {W.}~\bibnamefont
  {Liu}}, \bibinfo {author} {\bibfnamefont {X.}~\bibnamefont {Tan}}, \bibinfo
  {author} {\bibfnamefont {K.}~\bibnamefont {Yin}}, \bibinfo {author}
  {\bibfnamefont {H.}~\bibnamefont {Liu}}, \bibinfo {author} {\bibfnamefont
  {X.}~\bibnamefont {Tang}}, \bibinfo {author} {\bibfnamefont {J.}~\bibnamefont
  {Shi}}, \bibinfo {author} {\bibfnamefont {Q.}~\bibnamefont {Zhang}},\ and\
  \bibinfo {author} {\bibfnamefont {C.}~\bibnamefont {Uher}},\ }\bibfield
  {title} {\bibinfo {title} {Convergence of conduction bands as a means of
  enhancing thermoelectric performance of n-type mg 2 si 1- x sn x solid
  solutions},\ }\href@noop {} {\bibfield  {journal} {\bibinfo  {journal}
  {Physical review letters}\ }\textbf {\bibinfo {volume} {108}},\ \bibinfo
  {pages} {166601} (\bibinfo {year} {2012})}\BibitemShut {NoStop}%
\bibitem [{\citenamefont {Pei}\ \emph {et~al.}(2011{\natexlab{a}})\citenamefont
  {Pei}, \citenamefont {Shi}, \citenamefont {LaLonde}, \citenamefont {Wang},
  \citenamefont {Chen},\ and\ \citenamefont {Snyder}}]{pei2011convergence}%
  \BibitemOpen
  \bibfield  {author} {\bibinfo {author} {\bibfnamefont {Y.}~\bibnamefont
  {Pei}}, \bibinfo {author} {\bibfnamefont {X.}~\bibnamefont {Shi}}, \bibinfo
  {author} {\bibfnamefont {A.}~\bibnamefont {LaLonde}}, \bibinfo {author}
  {\bibfnamefont {H.}~\bibnamefont {Wang}}, \bibinfo {author} {\bibfnamefont
  {L.}~\bibnamefont {Chen}},\ and\ \bibinfo {author} {\bibfnamefont {G.~J.}\
  \bibnamefont {Snyder}},\ }\bibfield  {title} {\bibinfo {title} {Convergence
  of electronic bands for high performance bulk thermoelectrics},\ }\href@noop
  {} {\bibfield  {journal} {\bibinfo  {journal} {Nature}\ }\textbf {\bibinfo
  {volume} {473}},\ \bibinfo {pages} {66} (\bibinfo {year}
  {2011}{\natexlab{a}})}\BibitemShut {NoStop}%
\bibitem [{\citenamefont {Dresselhaus}\ \emph {et~al.}(2009)\citenamefont
  {Dresselhaus}, \citenamefont {Wakabayashi}, \citenamefont {Enoki},
  \citenamefont {Mori}, \citenamefont {Takai}, \citenamefont {Saito},
  \citenamefont {Murakami}, \citenamefont {Yamamoto},\ and\ \citenamefont
  {Sasaki}}]{dresselhaus2009kohn}%
  \BibitemOpen
  \bibfield  {author} {\bibinfo {author} {\bibfnamefont {M.}~\bibnamefont
  {Dresselhaus}}, \bibinfo {author} {\bibfnamefont {K.}~\bibnamefont
  {Wakabayashi}}, \bibinfo {author} {\bibfnamefont {T.}~\bibnamefont {Enoki}},
  \bibinfo {author} {\bibfnamefont {T.}~\bibnamefont {Mori}}, \bibinfo {author}
  {\bibfnamefont {K.}~\bibnamefont {Takai}}, \bibinfo {author} {\bibfnamefont
  {R.}~\bibnamefont {Saito}}, \bibinfo {author} {\bibfnamefont
  {S.}~\bibnamefont {Murakami}}, \bibinfo {author} {\bibfnamefont
  {M.}~\bibnamefont {Yamamoto}},\ and\ \bibinfo {author} {\bibfnamefont
  {K.-i.}\ \bibnamefont {Sasaki}},\ }\bibfield  {title} {\bibinfo {title} {Kohn
  anomalies in graphene nanoribbons},\ }\href@noop {} {\  (\bibinfo {year}
  {2009})}\BibitemShut {NoStop}%
\bibitem [{\citenamefont {Vineis}\ \emph {et~al.}(2010)\citenamefont {Vineis},
  \citenamefont {Shakouri}, \citenamefont {Majumdar},\ and\ \citenamefont
  {Kanatzidis}}]{vineis2010nanostructured}%
  \BibitemOpen
  \bibfield  {author} {\bibinfo {author} {\bibfnamefont {C.~J.}\ \bibnamefont
  {Vineis}}, \bibinfo {author} {\bibfnamefont {A.}~\bibnamefont {Shakouri}},
  \bibinfo {author} {\bibfnamefont {A.}~\bibnamefont {Majumdar}},\ and\
  \bibinfo {author} {\bibfnamefont {M.~G.}\ \bibnamefont {Kanatzidis}},\
  }\bibfield  {title} {\bibinfo {title} {Nanostructured thermoelectrics: big
  efficiency gains from small features},\ }\href@noop {} {\bibfield  {journal}
  {\bibinfo  {journal} {Advanced materials}\ }\textbf {\bibinfo {volume}
  {22}},\ \bibinfo {pages} {3970} (\bibinfo {year} {2010})}\BibitemShut
  {NoStop}%
\bibitem [{\citenamefont {Kanatzidis}(2010)}]{kanatzidis2010nanostructured}%
  \BibitemOpen
  \bibfield  {author} {\bibinfo {author} {\bibfnamefont {M.~G.}\ \bibnamefont
  {Kanatzidis}},\ }\bibfield  {title} {\bibinfo {title} {Nanostructured
  thermoelectrics: The new paradigm?},\ }\href@noop {} {\bibfield  {journal}
  {\bibinfo  {journal} {Chemistry of materials}\ }\textbf {\bibinfo {volume}
  {22}},\ \bibinfo {pages} {648} (\bibinfo {year} {2010})}\BibitemShut
  {NoStop}%
\bibitem [{\citenamefont {Biswas}\ \emph {et~al.}(2012)\citenamefont {Biswas},
  \citenamefont {He}, \citenamefont {Blum}, \citenamefont {Wu}, \citenamefont
  {Hogan}, \citenamefont {Seidman}, \citenamefont {Dravid},\ and\ \citenamefont
  {Kanatzidis}}]{biswas2012high}%
  \BibitemOpen
  \bibfield  {author} {\bibinfo {author} {\bibfnamefont {K.}~\bibnamefont
  {Biswas}}, \bibinfo {author} {\bibfnamefont {J.}~\bibnamefont {He}}, \bibinfo
  {author} {\bibfnamefont {I.~D.}\ \bibnamefont {Blum}}, \bibinfo {author}
  {\bibfnamefont {C.-I.}\ \bibnamefont {Wu}}, \bibinfo {author} {\bibfnamefont
  {T.~P.}\ \bibnamefont {Hogan}}, \bibinfo {author} {\bibfnamefont {D.~N.}\
  \bibnamefont {Seidman}}, \bibinfo {author} {\bibfnamefont {V.~P.}\
  \bibnamefont {Dravid}},\ and\ \bibinfo {author} {\bibfnamefont {M.~G.}\
  \bibnamefont {Kanatzidis}},\ }\bibfield  {title} {\bibinfo {title}
  {High-performance bulk thermoelectrics with all-scale hierarchical
  architectures},\ }\href@noop {} {\bibfield  {journal} {\bibinfo  {journal}
  {Nature}\ }\textbf {\bibinfo {volume} {489}},\ \bibinfo {pages} {414}
  (\bibinfo {year} {2012})}\BibitemShut {NoStop}%
\bibitem [{\citenamefont {Wang}\ \emph {et~al.}(2011)\citenamefont {Wang},
  \citenamefont {Alaniz}, \citenamefont {Jang}, \citenamefont {Garay},\ and\
  \citenamefont {Dames}}]{wang2011thermal}%
  \BibitemOpen
  \bibfield  {author} {\bibinfo {author} {\bibfnamefont {Z.}~\bibnamefont
  {Wang}}, \bibinfo {author} {\bibfnamefont {J.~E.}\ \bibnamefont {Alaniz}},
  \bibinfo {author} {\bibfnamefont {W.}~\bibnamefont {Jang}}, \bibinfo {author}
  {\bibfnamefont {J.~E.}\ \bibnamefont {Garay}},\ and\ \bibinfo {author}
  {\bibfnamefont {C.}~\bibnamefont {Dames}},\ }\bibfield  {title} {\bibinfo
  {title} {Thermal conductivity of nanocrystalline silicon: importance of grain
  size and frequency-dependent mean free paths},\ }\href@noop {} {\bibfield
  {journal} {\bibinfo  {journal} {Nano letters}\ }\textbf {\bibinfo {volume}
  {11}},\ \bibinfo {pages} {2206} (\bibinfo {year} {2011})}\BibitemShut
  {NoStop}%
\bibitem [{\citenamefont {Selli}\ \emph {et~al.}(2016)\citenamefont {Selli},
  \citenamefont {Boulfelfel}, \citenamefont {Schapotschnikow}, \citenamefont
  {Donadio},\ and\ \citenamefont {Leoni}}]{selli2016hierarchical}%
  \BibitemOpen
  \bibfield  {author} {\bibinfo {author} {\bibfnamefont {D.}~\bibnamefont
  {Selli}}, \bibinfo {author} {\bibfnamefont {S.~E.}\ \bibnamefont
  {Boulfelfel}}, \bibinfo {author} {\bibfnamefont {P.}~\bibnamefont
  {Schapotschnikow}}, \bibinfo {author} {\bibfnamefont {D.}~\bibnamefont
  {Donadio}},\ and\ \bibinfo {author} {\bibfnamefont {S.}~\bibnamefont
  {Leoni}},\ }\bibfield  {title} {\bibinfo {title} {Hierarchical
  thermoelectrics: crystal grain boundaries as scalable phonon scatterers},\
  }\href@noop {} {\bibfield  {journal} {\bibinfo  {journal} {Nanoscale}\
  }\textbf {\bibinfo {volume} {8}},\ \bibinfo {pages} {3729} (\bibinfo {year}
  {2016})}\BibitemShut {NoStop}%
\bibitem [{\citenamefont {Roychowdhury}\ \emph {et~al.}(2021)\citenamefont
  {Roychowdhury}, \citenamefont {Ghosh}, \citenamefont {Arora}, \citenamefont
  {Samanta}, \citenamefont {Xie}, \citenamefont {Singh}, \citenamefont {Soni},
  \citenamefont {He}, \citenamefont {Waghmare},\ and\ \citenamefont
  {Biswas}}]{roychowdhury2021enhanced}%
  \BibitemOpen
  \bibfield  {author} {\bibinfo {author} {\bibfnamefont {S.}~\bibnamefont
  {Roychowdhury}}, \bibinfo {author} {\bibfnamefont {T.}~\bibnamefont {Ghosh}},
  \bibinfo {author} {\bibfnamefont {R.}~\bibnamefont {Arora}}, \bibinfo
  {author} {\bibfnamefont {M.}~\bibnamefont {Samanta}}, \bibinfo {author}
  {\bibfnamefont {L.}~\bibnamefont {Xie}}, \bibinfo {author} {\bibfnamefont
  {N.~K.}\ \bibnamefont {Singh}}, \bibinfo {author} {\bibfnamefont
  {A.}~\bibnamefont {Soni}}, \bibinfo {author} {\bibfnamefont {J.}~\bibnamefont
  {He}}, \bibinfo {author} {\bibfnamefont {U.~V.}\ \bibnamefont {Waghmare}},\
  and\ \bibinfo {author} {\bibfnamefont {K.}~\bibnamefont {Biswas}},\
  }\bibfield  {title} {\bibinfo {title} {Enhanced atomic ordering leads to high
  thermoelectric performance in agsbte2},\ }\href@noop {} {\bibfield  {journal}
  {\bibinfo  {journal} {Science}\ }\textbf {\bibinfo {volume} {371}},\ \bibinfo
  {pages} {722} (\bibinfo {year} {2021})}\BibitemShut {NoStop}%
\bibitem [{\citenamefont {Rowe}(2018)}]{rowe2018crc}%
  \BibitemOpen
  \bibfield  {author} {\bibinfo {author} {\bibfnamefont {D.~M.}\ \bibnamefont
  {Rowe}},\ }\href@noop {} {\emph {\bibinfo {title} {CRC handbook of
  thermoelectrics}}}\ (\bibinfo  {publisher} {CRC press},\ \bibinfo {year}
  {2018})\BibitemShut {NoStop}%
\bibitem [{\citenamefont {Toberer}\ \emph {et~al.}(2010)\citenamefont
  {Toberer}, \citenamefont {May},\ and\ \citenamefont
  {Snyder}}]{toberer2010zintl}%
  \BibitemOpen
  \bibfield  {author} {\bibinfo {author} {\bibfnamefont {E.~S.}\ \bibnamefont
  {Toberer}}, \bibinfo {author} {\bibfnamefont {A.~F.}\ \bibnamefont {May}},\
  and\ \bibinfo {author} {\bibfnamefont {G.~J.}\ \bibnamefont {Snyder}},\
  }\bibfield  {title} {\bibinfo {title} {Zintl chemistry for designing high
  efficiency thermoelectric materials},\ }\href@noop {} {\bibfield  {journal}
  {\bibinfo  {journal} {Chemistry of Materials}\ }\textbf {\bibinfo {volume}
  {22}},\ \bibinfo {pages} {624} (\bibinfo {year} {2010})}\BibitemShut
  {NoStop}%
\bibitem [{\citenamefont {Shi}\ \emph {et~al.}(2011)\citenamefont {Shi},
  \citenamefont {Yang}, \citenamefont {Salvador}, \citenamefont {Chi},
  \citenamefont {Cho}, \citenamefont {Wang}, \citenamefont {Bai}, \citenamefont
  {Yang}, \citenamefont {Zhang},\ and\ \citenamefont {Chen}}]{shi2011multiple}%
  \BibitemOpen
  \bibfield  {author} {\bibinfo {author} {\bibfnamefont {X.}~\bibnamefont
  {Shi}}, \bibinfo {author} {\bibfnamefont {J.}~\bibnamefont {Yang}}, \bibinfo
  {author} {\bibfnamefont {J.~R.}\ \bibnamefont {Salvador}}, \bibinfo {author}
  {\bibfnamefont {M.}~\bibnamefont {Chi}}, \bibinfo {author} {\bibfnamefont
  {J.~Y.}\ \bibnamefont {Cho}}, \bibinfo {author} {\bibfnamefont
  {H.}~\bibnamefont {Wang}}, \bibinfo {author} {\bibfnamefont {S.}~\bibnamefont
  {Bai}}, \bibinfo {author} {\bibfnamefont {J.}~\bibnamefont {Yang}}, \bibinfo
  {author} {\bibfnamefont {W.}~\bibnamefont {Zhang}},\ and\ \bibinfo {author}
  {\bibfnamefont {L.}~\bibnamefont {Chen}},\ }\bibfield  {title} {\bibinfo
  {title} {Multiple-filled skutterudites: high thermoelectric figure of merit
  through separately optimizing electrical and thermal transports},\
  }\href@noop {} {\bibfield  {journal} {\bibinfo  {journal} {Journal of the
  American Chemical Society}\ }\textbf {\bibinfo {volume} {133}},\ \bibinfo
  {pages} {7837} (\bibinfo {year} {2011})}\BibitemShut {NoStop}%
\bibitem [{\citenamefont {Miyazaki}(2004)}]{miyazaki2004crystal}%
  \BibitemOpen
  \bibfield  {author} {\bibinfo {author} {\bibfnamefont {Y.}~\bibnamefont
  {Miyazaki}},\ }\bibfield  {title} {\bibinfo {title} {Crystal structure and
  thermoelectric properties of the misfit-layered cobalt oxides},\ }\href@noop
  {} {\bibfield  {journal} {\bibinfo  {journal} {Solid State Ionics}\ }\textbf
  {\bibinfo {volume} {172}},\ \bibinfo {pages} {463} (\bibinfo {year}
  {2004})}\BibitemShut {NoStop}%
\bibitem [{\citenamefont {He}\ \emph {et~al.}(2015)\citenamefont {He},
  \citenamefont {Zhang}, \citenamefont {Shi}, \citenamefont {Wei},\ and\
  \citenamefont {Chen}}]{he2015high}%
  \BibitemOpen
  \bibfield  {author} {\bibinfo {author} {\bibfnamefont {Y.}~\bibnamefont
  {He}}, \bibinfo {author} {\bibfnamefont {T.}~\bibnamefont {Zhang}}, \bibinfo
  {author} {\bibfnamefont {X.}~\bibnamefont {Shi}}, \bibinfo {author}
  {\bibfnamefont {S.-H.}\ \bibnamefont {Wei}},\ and\ \bibinfo {author}
  {\bibfnamefont {L.}~\bibnamefont {Chen}},\ }\bibfield  {title} {\bibinfo
  {title} {High thermoelectric performance in copper telluride},\ }\href@noop
  {} {\bibfield  {journal} {\bibinfo  {journal} {NPG Asia Materials}\ }\textbf
  {\bibinfo {volume} {7}},\ \bibinfo {pages} {e210} (\bibinfo {year}
  {2015})}\BibitemShut {NoStop}%
\bibitem [{\citenamefont {He}\ \emph {et~al.}(2014)\citenamefont {He},
  \citenamefont {Day}, \citenamefont {Zhang}, \citenamefont {Liu},
  \citenamefont {Shi}, \citenamefont {Chen},\ and\ \citenamefont
  {Snyder}}]{he2014high}%
  \BibitemOpen
  \bibfield  {author} {\bibinfo {author} {\bibfnamefont {Y.}~\bibnamefont
  {He}}, \bibinfo {author} {\bibfnamefont {T.}~\bibnamefont {Day}}, \bibinfo
  {author} {\bibfnamefont {T.}~\bibnamefont {Zhang}}, \bibinfo {author}
  {\bibfnamefont {H.}~\bibnamefont {Liu}}, \bibinfo {author} {\bibfnamefont
  {X.}~\bibnamefont {Shi}}, \bibinfo {author} {\bibfnamefont {L.}~\bibnamefont
  {Chen}},\ and\ \bibinfo {author} {\bibfnamefont {G.~J.}\ \bibnamefont
  {Snyder}},\ }\bibfield  {title} {\bibinfo {title} {High thermoelectric
  performance in non-toxic earth-abundant copper sulfide},\ }\href@noop {}
  {\bibfield  {journal} {\bibinfo  {journal} {Advanced Materials}\ }\textbf
  {\bibinfo {volume} {26}},\ \bibinfo {pages} {3974} (\bibinfo {year}
  {2014})}\BibitemShut {NoStop}%
\bibitem [{\citenamefont {Liu}\ \emph {et~al.}(2013)\citenamefont {Liu},
  \citenamefont {Yuan}, \citenamefont {Lu}, \citenamefont {Shi}, \citenamefont
  {Xu}, \citenamefont {He}, \citenamefont {Tang}, \citenamefont {Bai},
  \citenamefont {Zhang}, \citenamefont {Chen} \emph
  {et~al.}}]{liu2013ultrahigh}%
  \BibitemOpen
  \bibfield  {author} {\bibinfo {author} {\bibfnamefont {H.}~\bibnamefont
  {Liu}}, \bibinfo {author} {\bibfnamefont {X.}~\bibnamefont {Yuan}}, \bibinfo
  {author} {\bibfnamefont {P.}~\bibnamefont {Lu}}, \bibinfo {author}
  {\bibfnamefont {X.}~\bibnamefont {Shi}}, \bibinfo {author} {\bibfnamefont
  {F.}~\bibnamefont {Xu}}, \bibinfo {author} {\bibfnamefont {Y.}~\bibnamefont
  {He}}, \bibinfo {author} {\bibfnamefont {Y.}~\bibnamefont {Tang}}, \bibinfo
  {author} {\bibfnamefont {S.}~\bibnamefont {Bai}}, \bibinfo {author}
  {\bibfnamefont {W.}~\bibnamefont {Zhang}}, \bibinfo {author} {\bibfnamefont
  {L.}~\bibnamefont {Chen}}, \emph {et~al.},\ }\bibfield  {title} {\bibinfo
  {title} {Ultrahigh thermoelectric performance by electron and phonon critical
  scattering in cu2se1-xix},\ }\href@noop {} {\bibfield  {journal} {\bibinfo
  {journal} {Advanced Materials}\ }\textbf {\bibinfo {volume} {25}},\ \bibinfo
  {pages} {6607} (\bibinfo {year} {2013})}\BibitemShut {NoStop}%
\bibitem [{\citenamefont {Brown}\ \emph {et~al.}(2013)\citenamefont {Brown},
  \citenamefont {Day}, \citenamefont {Caillat},\ and\ \citenamefont
  {Snyder}}]{Brown2013}%
  \BibitemOpen
  \bibfield  {author} {\bibinfo {author} {\bibfnamefont {D.~R.}\ \bibnamefont
  {Brown}}, \bibinfo {author} {\bibfnamefont {T.}~\bibnamefont {Day}}, \bibinfo
  {author} {\bibfnamefont {T.}~\bibnamefont {Caillat}},\ and\ \bibinfo {author}
  {\bibfnamefont {G.~J.}\ \bibnamefont {Snyder}},\ }\bibfield  {title}
  {\bibinfo {title} {Chemical stability of (ag,cu)$_2$se: a historical
  overview},\ }\href {https://doi.org/10.1007/s11664-013-2506-2} {\bibfield
  {journal} {\bibinfo  {journal} {Journal of Electronic Materials}\ }\textbf
  {\bibinfo {volume} {42}},\ \bibinfo {pages} {2014} (\bibinfo {year}
  {2013})}\BibitemShut {NoStop}%
\bibitem [{\citenamefont {Dennler}\ \emph {et~al.}(2014)\citenamefont
  {Dennler}, \citenamefont {Chmielowski}, \citenamefont {Jacob}, \citenamefont
  {Capet}, \citenamefont {Roussel}, \citenamefont {Zastrow}, \citenamefont
  {Nielsch}, \citenamefont {Opahle},\ and\ \citenamefont
  {Madsen}}]{dennler2014binary}%
  \BibitemOpen
  \bibfield  {author} {\bibinfo {author} {\bibfnamefont {G.}~\bibnamefont
  {Dennler}}, \bibinfo {author} {\bibfnamefont {R.}~\bibnamefont
  {Chmielowski}}, \bibinfo {author} {\bibfnamefont {S.}~\bibnamefont {Jacob}},
  \bibinfo {author} {\bibfnamefont {F.}~\bibnamefont {Capet}}, \bibinfo
  {author} {\bibfnamefont {P.}~\bibnamefont {Roussel}}, \bibinfo {author}
  {\bibfnamefont {S.}~\bibnamefont {Zastrow}}, \bibinfo {author} {\bibfnamefont
  {K.}~\bibnamefont {Nielsch}}, \bibinfo {author} {\bibfnamefont
  {I.}~\bibnamefont {Opahle}},\ and\ \bibinfo {author} {\bibfnamefont {G.~K.}\
  \bibnamefont {Madsen}},\ }\bibfield  {title} {\bibinfo {title} {Are binary
  copper sulfides/selenides really new and promising thermoelectric
  materials?},\ }\href@noop {} {\bibfield  {journal} {\bibinfo  {journal}
  {Advanced Energy Materials}\ }\textbf {\bibinfo {volume} {4}},\ \bibinfo
  {pages} {1301581} (\bibinfo {year} {2014})}\BibitemShut {NoStop}%
\bibitem [{\citenamefont {Mao}\ \emph {et~al.}(2020)\citenamefont {Mao},
  \citenamefont {Qiu}, \citenamefont {Hu}, \citenamefont {Du}, \citenamefont
  {Zhao}, \citenamefont {Wei}, \citenamefont {Xiao}, \citenamefont {Shi},\ and\
  \citenamefont {Chen}}]{TaoMao2020}%
  \BibitemOpen
  \bibfield  {author} {\bibinfo {author} {\bibfnamefont {T.}~\bibnamefont
  {Mao}}, \bibinfo {author} {\bibfnamefont {P.}~\bibnamefont {Qiu}}, \bibinfo
  {author} {\bibfnamefont {P.}~\bibnamefont {Hu}}, \bibinfo {author}
  {\bibfnamefont {X.}~\bibnamefont {Du}}, \bibinfo {author} {\bibfnamefont
  {K.}~\bibnamefont {Zhao}}, \bibinfo {author} {\bibfnamefont {T.-R.}\
  \bibnamefont {Wei}}, \bibinfo {author} {\bibfnamefont {J.}~\bibnamefont
  {Xiao}}, \bibinfo {author} {\bibfnamefont {X.}~\bibnamefont {Shi}},\ and\
  \bibinfo {author} {\bibfnamefont {L.}~\bibnamefont {Chen}},\ }\bibfield
  {title} {\bibinfo {title} {Decoupling thermoelectric performance and
  stability in liquid-like thermoelectric materials},\ }\href@noop {}
  {\bibfield  {journal} {\bibinfo  {journal} {Advanced Science}\ }\textbf
  {\bibinfo {volume} {7}},\ \bibinfo {pages} {1901598} (\bibinfo {year}
  {2020})}\BibitemShut {NoStop}%
\bibitem [{\citenamefont {Yang}\ \emph {et~al.}(2020)\citenamefont {Yang},
  \citenamefont {Su}, \citenamefont {Li}, \citenamefont {Bai}, \citenamefont
  {Wang}, \citenamefont {Li}, \citenamefont {Tang}, \citenamefont {Tang},
  \citenamefont {Luo}, \citenamefont {Yan}, \citenamefont {Wu}, \citenamefont
  {Yang}, \citenamefont {Zhang}, \citenamefont {Uher}, \citenamefont
  {Kanatzidis},\ and\ \citenamefont {Tang}}]{Dongwang2020}%
  \BibitemOpen
  \bibfield  {author} {\bibinfo {author} {\bibfnamefont {D.}~\bibnamefont
  {Yang}}, \bibinfo {author} {\bibfnamefont {X.}~\bibnamefont {Su}}, \bibinfo
  {author} {\bibfnamefont {J.}~\bibnamefont {Li}}, \bibinfo {author}
  {\bibfnamefont {H.}~\bibnamefont {Bai}}, \bibinfo {author} {\bibfnamefont
  {S.}~\bibnamefont {Wang}}, \bibinfo {author} {\bibfnamefont {Z.}~\bibnamefont
  {Li}}, \bibinfo {author} {\bibfnamefont {H.}~\bibnamefont {Tang}}, \bibinfo
  {author} {\bibfnamefont {K.}~\bibnamefont {Tang}}, \bibinfo {author}
  {\bibfnamefont {T.}~\bibnamefont {Luo}}, \bibinfo {author} {\bibfnamefont
  {Y.}~\bibnamefont {Yan}}, \bibinfo {author} {\bibfnamefont {J.}~\bibnamefont
  {Wu}}, \bibinfo {author} {\bibfnamefont {J.}~\bibnamefont {Yang}}, \bibinfo
  {author} {\bibfnamefont {Q.}~\bibnamefont {Zhang}}, \bibinfo {author}
  {\bibfnamefont {C.}~\bibnamefont {Uher}}, \bibinfo {author} {\bibfnamefont
  {M.~G.}\ \bibnamefont {Kanatzidis}},\ and\ \bibinfo {author} {\bibfnamefont
  {X.}~\bibnamefont {Tang}},\ }\bibfield  {title} {\bibinfo {title} {Blocking
  ion migration stabilizes the high thermoelectric performance in cu2se
  composites},\ }\href@noop {} {\bibfield  {journal} {\bibinfo  {journal}
  {Advanced Materials}\ }\textbf {\bibinfo {volume} {32}},\ \bibinfo {pages}
  {2003730} (\bibinfo {year} {2020})}\BibitemShut {NoStop}%
\bibitem [{\citenamefont {Zhao}\ \emph {et~al.}(2022)\citenamefont {Zhao},
  \citenamefont {Zhu}, \citenamefont {Zhu}, \citenamefont {Chen}, \citenamefont
  {Lei}, \citenamefont {Ren}, \citenamefont {Wei}, \citenamefont {Qiu},
  \citenamefont {Xu}, \citenamefont {Chen}, \citenamefont {He},\ and\
  \citenamefont {Shi}}]{Kunpeng2022}%
  \BibitemOpen
  \bibfield  {author} {\bibinfo {author} {\bibfnamefont {K.}~\bibnamefont
  {Zhao}}, \bibinfo {author} {\bibfnamefont {C.}~\bibnamefont {Zhu}}, \bibinfo
  {author} {\bibfnamefont {M.}~\bibnamefont {Zhu}}, \bibinfo {author}
  {\bibfnamefont {H.}~\bibnamefont {Chen}}, \bibinfo {author} {\bibfnamefont
  {J.}~\bibnamefont {Lei}}, \bibinfo {author} {\bibfnamefont {Q.}~\bibnamefont
  {Ren}}, \bibinfo {author} {\bibfnamefont {T.-R.}\ \bibnamefont {Wei}},
  \bibinfo {author} {\bibfnamefont {P.}~\bibnamefont {Qiu}}, \bibinfo {author}
  {\bibfnamefont {F.}~\bibnamefont {Xu}}, \bibinfo {author} {\bibfnamefont
  {L.}~\bibnamefont {Chen}}, \bibinfo {author} {\bibfnamefont {J.}~\bibnamefont
  {He}},\ and\ \bibinfo {author} {\bibfnamefont {X.}~\bibnamefont {Shi}},\
  }\bibfield  {title} {\bibinfo {title} {Structural modularization of cu2te
  leading to high thermoelectric performance near the mott–ioffe–regel
  limit},\ }\href {https://doi.org/https://doi.org/10.1002/adma.202108573}
  {\bibfield  {journal} {\bibinfo  {journal} {Advanced Materials}\ }\textbf
  {\bibinfo {volume} {34}},\ \bibinfo {pages} {2108573} (\bibinfo {year}
  {2022})}\BibitemShut {NoStop}%
\bibitem [{\citenamefont {Capps}\ \emph {et~al.}(2010)\citenamefont {Capps},
  \citenamefont {Drymiotis}, \citenamefont {Lindsey},\ and\ \citenamefont
  {Tritt}}]{capps2010significant}%
  \BibitemOpen
  \bibfield  {author} {\bibinfo {author} {\bibfnamefont {J.}~\bibnamefont
  {Capps}}, \bibinfo {author} {\bibfnamefont {F.}~\bibnamefont {Drymiotis}},
  \bibinfo {author} {\bibfnamefont {S.}~\bibnamefont {Lindsey}},\ and\ \bibinfo
  {author} {\bibfnamefont {T.}~\bibnamefont {Tritt}},\ }\bibfield  {title}
  {\bibinfo {title} {Significant enhancement of the dimensionless
  thermoelectric figure of merit of the binary ag2te},\ }\href@noop {}
  {\bibfield  {journal} {\bibinfo  {journal} {Philosophical magazine letters}\
  }\textbf {\bibinfo {volume} {90}},\ \bibinfo {pages} {677} (\bibinfo {year}
  {2010})}\BibitemShut {NoStop}%
\bibitem [{\citenamefont {Zhu}\ \emph {et~al.}(2015)\citenamefont {Zhu},
  \citenamefont {Luo}, \citenamefont {Zhao},\ and\ \citenamefont
  {Liang}}]{zhu2015enhanced}%
  \BibitemOpen
  \bibfield  {author} {\bibinfo {author} {\bibfnamefont {H.}~\bibnamefont
  {Zhu}}, \bibinfo {author} {\bibfnamefont {J.}~\bibnamefont {Luo}}, \bibinfo
  {author} {\bibfnamefont {H.}~\bibnamefont {Zhao}},\ and\ \bibinfo {author}
  {\bibfnamefont {J.}~\bibnamefont {Liang}},\ }\bibfield  {title} {\bibinfo
  {title} {Enhanced thermoelectric properties of p-type ag 2 te by cu
  substitution},\ }\href@noop {} {\bibfield  {journal} {\bibinfo  {journal}
  {Journal of Materials Chemistry A}\ }\textbf {\bibinfo {volume} {3}},\
  \bibinfo {pages} {10303} (\bibinfo {year} {2015})}\BibitemShut {NoStop}%
\bibitem [{\citenamefont {Taylor}\ and\ \citenamefont
  {Wood}(1961)}]{taylor1961thermoelectric}%
  \BibitemOpen
  \bibfield  {author} {\bibinfo {author} {\bibfnamefont {P.}~\bibnamefont
  {Taylor}}\ and\ \bibinfo {author} {\bibfnamefont {C.}~\bibnamefont {Wood}},\
  }\bibfield  {title} {\bibinfo {title} {Thermoelectric properties of ag2te},\
  }\href@noop {} {\bibfield  {journal} {\bibinfo  {journal} {Journal of Applied
  Physics}\ }\textbf {\bibinfo {volume} {32}},\ \bibinfo {pages} {1} (\bibinfo
  {year} {1961})}\BibitemShut {NoStop}%
\bibitem [{\citenamefont {Jahangirli}\ \emph {et~al.}(2018)\citenamefont
  {Jahangirli}, \citenamefont {Alekperov},\ and\ \citenamefont
  {Eyyubov}}]{jahangirli2018ab}%
  \BibitemOpen
  \bibfield  {author} {\bibinfo {author} {\bibfnamefont {Z.}~\bibnamefont
  {Jahangirli}}, \bibinfo {author} {\bibfnamefont {O.}~\bibnamefont
  {Alekperov}},\ and\ \bibinfo {author} {\bibfnamefont {Q.}~\bibnamefont
  {Eyyubov}},\ }\bibfield  {title} {\bibinfo {title} {Ab-initio investigation
  of the electronic structure, optical properties, and lattice dynamics of
  $\beta$-ag2te},\ }\href@noop {} {\bibfield  {journal} {\bibinfo  {journal}
  {physica status solidi (b)}\ }\textbf {\bibinfo {volume} {255}},\ \bibinfo
  {pages} {1800344} (\bibinfo {year} {2018})}\BibitemShut {NoStop}%
\bibitem [{\citenamefont {Pei}\ \emph {et~al.}(2011{\natexlab{b}})\citenamefont
  {Pei}, \citenamefont {Heinz},\ and\ \citenamefont
  {Snyder}}]{pei2011alloying}%
  \BibitemOpen
  \bibfield  {author} {\bibinfo {author} {\bibfnamefont {Y.}~\bibnamefont
  {Pei}}, \bibinfo {author} {\bibfnamefont {N.~A.}\ \bibnamefont {Heinz}},\
  and\ \bibinfo {author} {\bibfnamefont {G.~J.}\ \bibnamefont {Snyder}},\
  }\bibfield  {title} {\bibinfo {title} {Alloying to increase the band gap for
  improving thermoelectric properties of ag 2 te},\ }\href@noop {} {\bibfield
  {journal} {\bibinfo  {journal} {Journal of Materials Chemistry}\ }\textbf
  {\bibinfo {volume} {21}},\ \bibinfo {pages} {18256} (\bibinfo {year}
  {2011}{\natexlab{b}})}\BibitemShut {NoStop}%
\bibitem [{\citenamefont {Bailey}\ and\ \citenamefont
  {Uher}(2017)}]{bailey2017potential}%
  \BibitemOpen
  \bibfield  {author} {\bibinfo {author} {\bibfnamefont {T.~P.}\ \bibnamefont
  {Bailey}}\ and\ \bibinfo {author} {\bibfnamefont {C.}~\bibnamefont {Uher}},\
  }\bibfield  {title} {\bibinfo {title} {Potential for superionic conductors in
  thermoelectric applications},\ }\href@noop {} {\bibfield  {journal} {\bibinfo
   {journal} {Current Opinion in Green and Sustainable Chemistry}\ }\textbf
  {\bibinfo {volume} {4}},\ \bibinfo {pages} {58} (\bibinfo {year}
  {2017})}\BibitemShut {NoStop}%
\bibitem [{\citenamefont {Cadavid}\ \emph {et~al.}(2013)\citenamefont
  {Cadavid}, \citenamefont {Ib}, \citenamefont {Shavel}, \citenamefont {Dura},
  \citenamefont {De~La~Torre},\ and\ \citenamefont
  {Cabot}}]{cadavid2013organic}%
  \BibitemOpen
  \bibfield  {author} {\bibinfo {author} {\bibfnamefont {D.}~\bibnamefont
  {Cadavid}}, \bibinfo {author} {\bibfnamefont {M.}~\bibnamefont {Ib}},
  \bibinfo {author} {\bibfnamefont {A.}~\bibnamefont {Shavel}}, \bibinfo
  {author} {\bibfnamefont {O.~J.}\ \bibnamefont {Dura}}, \bibinfo {author}
  {\bibfnamefont {M.~L.}\ \bibnamefont {De~La~Torre}},\ and\ \bibinfo {author}
  {\bibfnamefont {A.}~\bibnamefont {Cabot}},\ }\bibfield  {title} {\bibinfo
  {title} {Organic ligand displacement by metal salts to enhance nanoparticle
  functionality: thermoelectric properties of ag 2 te},\ }\href@noop {}
  {\bibfield  {journal} {\bibinfo  {journal} {Journal of Materials Chemistry
  A}\ }\textbf {\bibinfo {volume} {1}},\ \bibinfo {pages} {4864} (\bibinfo
  {year} {2013})}\BibitemShut {NoStop}%
\bibitem [{\citenamefont {Aliev}(2003)}]{aliev2003electrical}%
  \BibitemOpen
  \bibfield  {author} {\bibinfo {author} {\bibfnamefont {F.}~\bibnamefont
  {Aliev}},\ }\bibfield  {title} {\bibinfo {title} {Electrical and
  thermoelectric properties of p-ag 2 te in the $\beta$ phase},\ }\href@noop {}
  {\bibfield  {journal} {\bibinfo  {journal} {Semiconductors}\ }\textbf
  {\bibinfo {volume} {37}},\ \bibinfo {pages} {1057} (\bibinfo {year}
  {2003})}\BibitemShut {NoStop}%
\bibitem [{\citenamefont {Dalven}\ and\ \citenamefont
  {Gill}(1966)}]{dalven1966energy}%
  \BibitemOpen
  \bibfield  {author} {\bibinfo {author} {\bibfnamefont {R.}~\bibnamefont
  {Dalven}}\ and\ \bibinfo {author} {\bibfnamefont {R.}~\bibnamefont {Gill}},\
  }\bibfield  {title} {\bibinfo {title} {Energy gap in $\beta$- ag 2 te},\
  }\href@noop {} {\bibfield  {journal} {\bibinfo  {journal} {Physical Review}\
  }\textbf {\bibinfo {volume} {143}},\ \bibinfo {pages} {666} (\bibinfo {year}
  {1966})}\BibitemShut {NoStop}%
\bibitem [{\citenamefont {Wood}\ \emph {et~al.}(1961)\citenamefont {Wood},
  \citenamefont {Harrap},\ and\ \citenamefont {Kane}}]{wood1961degeneracy}%
  \BibitemOpen
  \bibfield  {author} {\bibinfo {author} {\bibfnamefont {C.}~\bibnamefont
  {Wood}}, \bibinfo {author} {\bibfnamefont {V.}~\bibnamefont {Harrap}},\ and\
  \bibinfo {author} {\bibfnamefont {W.}~\bibnamefont {Kane}},\ }\bibfield
  {title} {\bibinfo {title} {Degeneracy in ag 2 te},\ }\href@noop {} {\bibfield
   {journal} {\bibinfo  {journal} {Physical Review}\ }\textbf {\bibinfo
  {volume} {121}},\ \bibinfo {pages} {978} (\bibinfo {year}
  {1961})}\BibitemShut {NoStop}%
\bibitem [{\citenamefont {Chang}\ \emph {et~al.}(2019)\citenamefont {Chang},
  \citenamefont {Guo}, \citenamefont {Tang}, \citenamefont {Zhang},
  \citenamefont {Feng},\ and\ \citenamefont {Ge}}]{chang2019facile}%
  \BibitemOpen
  \bibfield  {author} {\bibinfo {author} {\bibfnamefont {Y.}~\bibnamefont
  {Chang}}, \bibinfo {author} {\bibfnamefont {J.}~\bibnamefont {Guo}}, \bibinfo
  {author} {\bibfnamefont {Y.-Q.}\ \bibnamefont {Tang}}, \bibinfo {author}
  {\bibfnamefont {Y.-X.}\ \bibnamefont {Zhang}}, \bibinfo {author}
  {\bibfnamefont {J.}~\bibnamefont {Feng}},\ and\ \bibinfo {author}
  {\bibfnamefont {Z.-H.}\ \bibnamefont {Ge}},\ }\bibfield  {title} {\bibinfo
  {title} {Facile synthesis of ag 2 te nanowires and thermoelectric properties
  of ag 2 te polycrystals sintered by spark plasma sintering},\ }\href@noop {}
  {\bibfield  {journal} {\bibinfo  {journal} {CrystEngComm}\ }\textbf {\bibinfo
  {volume} {21}},\ \bibinfo {pages} {1718} (\bibinfo {year}
  {2019})}\BibitemShut {NoStop}%
\bibitem [{\citenamefont {Jung}\ \emph {et~al.}(2012)\citenamefont {Jung},
  \citenamefont {Kurosaki}, \citenamefont {Ohishi}, \citenamefont {Muta},\ and\
  \citenamefont {Yamanaka}}]{jung2012effect}%
  \BibitemOpen
  \bibfield  {author} {\bibinfo {author} {\bibfnamefont {D.-y.}\ \bibnamefont
  {Jung}}, \bibinfo {author} {\bibfnamefont {K.}~\bibnamefont {Kurosaki}},
  \bibinfo {author} {\bibfnamefont {Y.}~\bibnamefont {Ohishi}}, \bibinfo
  {author} {\bibfnamefont {H.}~\bibnamefont {Muta}},\ and\ \bibinfo {author}
  {\bibfnamefont {S.}~\bibnamefont {Yamanaka}},\ }\bibfield  {title} {\bibinfo
  {title} {Effect of phase transition on the thermoelectric properties of
  ag2te},\ }\href@noop {} {\bibfield  {journal} {\bibinfo  {journal} {Materials
  Transactions}\ ,\ \bibinfo {pages} {E}} (\bibinfo {year} {2012})}\BibitemShut
  {NoStop}%
\bibitem [{\citenamefont {Fujikane}\ \emph {et~al.}(2005)\citenamefont
  {Fujikane}, \citenamefont {Kurosaki}, \citenamefont {Muta},\ and\
  \citenamefont {Yamanaka}}]{fujikane2005thermoelectric}%
  \BibitemOpen
  \bibfield  {author} {\bibinfo {author} {\bibfnamefont {M.}~\bibnamefont
  {Fujikane}}, \bibinfo {author} {\bibfnamefont {K.}~\bibnamefont {Kurosaki}},
  \bibinfo {author} {\bibfnamefont {H.}~\bibnamefont {Muta}},\ and\ \bibinfo
  {author} {\bibfnamefont {S.}~\bibnamefont {Yamanaka}},\ }\bibfield  {title}
  {\bibinfo {title} {Thermoelectric properties of $\alpha$-and $\beta$-ag2te},\
  }\href@noop {} {\bibfield  {journal} {\bibinfo  {journal} {Journal of alloys
  and compounds}\ }\textbf {\bibinfo {volume} {393}},\ \bibinfo {pages} {299}
  (\bibinfo {year} {2005})}\BibitemShut {NoStop}%
\bibitem [{\citenamefont {Aliev}\ and\ \citenamefont
  {Eminova}(2015)}]{aliev2015dependence}%
  \BibitemOpen
  \bibfield  {author} {\bibinfo {author} {\bibfnamefont {F.}~\bibnamefont
  {Aliev}}\ and\ \bibinfo {author} {\bibfnamefont {V.}~\bibnamefont
  {Eminova}},\ }\bibfield  {title} {\bibinfo {title} {Dependence of the spectra
  of charge carriers on the concentration of defects in silver telluride},\
  }\href@noop {} {\bibfield  {journal} {\bibinfo  {journal} {Physics of the
  Solid State}\ }\textbf {\bibinfo {volume} {57}},\ \bibinfo {pages} {1325}
  (\bibinfo {year} {2015})}\BibitemShut {NoStop}%
\bibitem [{\citenamefont {Shi}\ \emph {et~al.}(2018)\citenamefont {Shi},
  \citenamefont {Chen}, \citenamefont {Hao}, \citenamefont {Liu}, \citenamefont
  {Wang}, \citenamefont {Qiu}, \citenamefont {Burkhardt}, \citenamefont
  {Grin},\ and\ \citenamefont {Chen}}]{shi2018room}%
  \BibitemOpen
  \bibfield  {author} {\bibinfo {author} {\bibfnamefont {X.}~\bibnamefont
  {Shi}}, \bibinfo {author} {\bibfnamefont {H.}~\bibnamefont {Chen}}, \bibinfo
  {author} {\bibfnamefont {F.}~\bibnamefont {Hao}}, \bibinfo {author}
  {\bibfnamefont {R.}~\bibnamefont {Liu}}, \bibinfo {author} {\bibfnamefont
  {T.}~\bibnamefont {Wang}}, \bibinfo {author} {\bibfnamefont {P.}~\bibnamefont
  {Qiu}}, \bibinfo {author} {\bibfnamefont {U.}~\bibnamefont {Burkhardt}},
  \bibinfo {author} {\bibfnamefont {Y.}~\bibnamefont {Grin}},\ and\ \bibinfo
  {author} {\bibfnamefont {L.}~\bibnamefont {Chen}},\ }\bibfield  {title}
  {\bibinfo {title} {Room-temperature ductile inorganic semiconductor},\
  }\href@noop {} {\bibfield  {journal} {\bibinfo  {journal} {Nature Materials}\
  }\textbf {\bibinfo {volume} {17}},\ \bibinfo {pages} {421} (\bibinfo {year}
  {2018})}\BibitemShut {NoStop}%
\bibitem [{\citenamefont {Malik}\ \emph {et~al.}(2020)\citenamefont {Malik},
  \citenamefont {Srivastava}, \citenamefont {Surthi}, \citenamefont {Gayner},\
  and\ \citenamefont {Kar}}]{malik2020enhanced}%
  \BibitemOpen
  \bibfield  {author} {\bibinfo {author} {\bibfnamefont {I.}~\bibnamefont
  {Malik}}, \bibinfo {author} {\bibfnamefont {T.}~\bibnamefont {Srivastava}},
  \bibinfo {author} {\bibfnamefont {K.~K.}\ \bibnamefont {Surthi}}, \bibinfo
  {author} {\bibfnamefont {C.}~\bibnamefont {Gayner}},\ and\ \bibinfo {author}
  {\bibfnamefont {K.~K.}\ \bibnamefont {Kar}},\ }\bibfield  {title} {\bibinfo
  {title} {Enhanced thermoelectric performance of n-type bi2te3 alloyed with
  low cost and highly abundant sulfur},\ }\href@noop {} {\bibfield  {journal}
  {\bibinfo  {journal} {Materials Chemistry and Physics}\ }\textbf {\bibinfo
  {volume} {255}},\ \bibinfo {pages} {123598} (\bibinfo {year}
  {2020})}\BibitemShut {NoStop}%
\bibitem [{\citenamefont {Yang}\ \emph {et~al.}(2014)\citenamefont {Yang},
  \citenamefont {Bahk}, \citenamefont {Day}, \citenamefont {Mohammed},
  \citenamefont {Min}, \citenamefont {Snyder}, \citenamefont {Shakouri},\ and\
  \citenamefont {Wu}}]{yang2014composition}%
  \BibitemOpen
  \bibfield  {author} {\bibinfo {author} {\bibfnamefont {H.}~\bibnamefont
  {Yang}}, \bibinfo {author} {\bibfnamefont {J.-H.}\ \bibnamefont {Bahk}},
  \bibinfo {author} {\bibfnamefont {T.}~\bibnamefont {Day}}, \bibinfo {author}
  {\bibfnamefont {A.~M.}\ \bibnamefont {Mohammed}}, \bibinfo {author}
  {\bibfnamefont {B.}~\bibnamefont {Min}}, \bibinfo {author} {\bibfnamefont
  {G.~J.}\ \bibnamefont {Snyder}}, \bibinfo {author} {\bibfnamefont
  {A.}~\bibnamefont {Shakouri}},\ and\ \bibinfo {author} {\bibfnamefont
  {Y.}~\bibnamefont {Wu}},\ }\bibfield  {title} {\bibinfo {title} {Composition
  modulation of ag2te nanowires for tunable electrical and thermal
  properties},\ }\href@noop {} {\bibfield  {journal} {\bibinfo  {journal} {Nano
  letters}\ }\textbf {\bibinfo {volume} {14}},\ \bibinfo {pages} {5398}
  (\bibinfo {year} {2014})}\BibitemShut {NoStop}%
\bibitem [{\citenamefont {Cadavid}\ \emph {et~al.}(2012)\citenamefont
  {Cadavid}, \citenamefont {Ibanez}, \citenamefont {Gorsse}, \citenamefont
  {Lopez}, \citenamefont {Cirera}, \citenamefont {Morante},\ and\ \citenamefont
  {Cabot}}]{cadavid2012bottom}%
  \BibitemOpen
  \bibfield  {author} {\bibinfo {author} {\bibfnamefont {D.}~\bibnamefont
  {Cadavid}}, \bibinfo {author} {\bibfnamefont {M.}~\bibnamefont {Ibanez}},
  \bibinfo {author} {\bibfnamefont {S.}~\bibnamefont {Gorsse}}, \bibinfo
  {author} {\bibfnamefont {A.~M.}\ \bibnamefont {Lopez}}, \bibinfo {author}
  {\bibfnamefont {A.}~\bibnamefont {Cirera}}, \bibinfo {author} {\bibfnamefont
  {J.~R.}\ \bibnamefont {Morante}},\ and\ \bibinfo {author} {\bibfnamefont
  {A.}~\bibnamefont {Cabot}},\ }\bibfield  {title} {\bibinfo {title} {Bottom-up
  processing of thermoelectric nanocomposites from colloidal nanocrystal
  building blocks: the case of ag2te--pbte},\ }\href@noop {} {\bibfield
  {journal} {\bibinfo  {journal} {Journal of nanoparticle research}\ }\textbf
  {\bibinfo {volume} {14}},\ \bibinfo {pages} {1} (\bibinfo {year}
  {2012})}\BibitemShut {NoStop}%
\bibitem [{\citenamefont {Hu}\ \emph {et~al.}(2021)\citenamefont {Hu},
  \citenamefont {Xia}, \citenamefont {Wang}, \citenamefont {Fu}, \citenamefont
  {Zhu},\ and\ \citenamefont {Zhao}}]{hu2021fast}%
  \BibitemOpen
  \bibfield  {author} {\bibinfo {author} {\bibfnamefont {H.}~\bibnamefont
  {Hu}}, \bibinfo {author} {\bibfnamefont {K.}~\bibnamefont {Xia}}, \bibinfo
  {author} {\bibfnamefont {Y.}~\bibnamefont {Wang}}, \bibinfo {author}
  {\bibfnamefont {C.}~\bibnamefont {Fu}}, \bibinfo {author} {\bibfnamefont
  {T.}~\bibnamefont {Zhu}},\ and\ \bibinfo {author} {\bibfnamefont
  {X.}~\bibnamefont {Zhao}},\ }\bibfield  {title} {\bibinfo {title} {Fast
  synthesis and improved electrical stability in n-type ag2te thermoelectric
  materials},\ }\href@noop {} {\bibfield  {journal} {\bibinfo  {journal}
  {Journal of Materials Science \& Technology}\ }\textbf {\bibinfo {volume}
  {91}},\ \bibinfo {pages} {241} (\bibinfo {year} {2021})}\BibitemShut
  {NoStop}%
\bibitem [{\citenamefont {Zhou}\ \emph {et~al.}(2012)\citenamefont {Zhou},
  \citenamefont {Zhao}, \citenamefont {Lu}, \citenamefont {Zhu}, \citenamefont
  {Fan}, \citenamefont {Ma}, \citenamefont {Hng},\ and\ \citenamefont
  {Yan}}]{zhou2012preparation}%
  \BibitemOpen
  \bibfield  {author} {\bibinfo {author} {\bibfnamefont {W.}~\bibnamefont
  {Zhou}}, \bibinfo {author} {\bibfnamefont {W.}~\bibnamefont {Zhao}}, \bibinfo
  {author} {\bibfnamefont {Z.}~\bibnamefont {Lu}}, \bibinfo {author}
  {\bibfnamefont {J.}~\bibnamefont {Zhu}}, \bibinfo {author} {\bibfnamefont
  {S.}~\bibnamefont {Fan}}, \bibinfo {author} {\bibfnamefont {J.}~\bibnamefont
  {Ma}}, \bibinfo {author} {\bibfnamefont {H.~H.}\ \bibnamefont {Hng}},\ and\
  \bibinfo {author} {\bibfnamefont {Q.}~\bibnamefont {Yan}},\ }\bibfield
  {title} {\bibinfo {title} {Preparation and thermoelectric properties of
  sulfur doped ag 2 te nanoparticles via solvothermal methods},\ }\href@noop {}
  {\bibfield  {journal} {\bibinfo  {journal} {Nanoscale}\ }\textbf {\bibinfo
  {volume} {4}},\ \bibinfo {pages} {3926} (\bibinfo {year} {2012})}\BibitemShut
  {NoStop}%
\bibitem [{\citenamefont {Van~der Lee}\ and\ \citenamefont
  {De~Boer}(1993)}]{van1993redetermination}%
  \BibitemOpen
  \bibfield  {author} {\bibinfo {author} {\bibfnamefont {A.}~\bibnamefont
  {Van~der Lee}}\ and\ \bibinfo {author} {\bibfnamefont {J.}~\bibnamefont
  {De~Boer}},\ }\bibfield  {title} {\bibinfo {title} {Redetermination of the
  structure of hessite, ag2te-iii},\ }\href@noop {} {\bibfield  {journal}
  {\bibinfo  {journal} {Acta Crystallographica Section C: Crystal Structure
  Communications}\ }\textbf {\bibinfo {volume} {49}},\ \bibinfo {pages} {1444}
  (\bibinfo {year} {1993})}\BibitemShut {NoStop}%
\bibitem [{\citenamefont {Yokoyama}\ \emph {et~al.}(1996)\citenamefont
  {Yokoyama}, \citenamefont {Tamura}, \citenamefont {Usui},\ and\ \citenamefont
  {Jimbo}}]{yokoyama1996simulation}%
  \BibitemOpen
  \bibfield  {author} {\bibinfo {author} {\bibfnamefont {T.}~\bibnamefont
  {Yokoyama}}, \bibinfo {author} {\bibfnamefont {K.}~\bibnamefont {Tamura}},
  \bibinfo {author} {\bibfnamefont {H.}~\bibnamefont {Usui}},\ and\ \bibinfo
  {author} {\bibfnamefont {G.}~\bibnamefont {Jimbo}},\ }\bibfield  {title}
  {\bibinfo {title} {Simulation of ball behavior in a vibration mill in
  relation with its grinding rate: effects of fractional ball filling and
  liquid viscosity},\ }\href@noop {} {\bibfield  {journal} {\bibinfo  {journal}
  {International Journal of Mineral Processing}\ }\textbf {\bibinfo {volume}
  {44}},\ \bibinfo {pages} {413} (\bibinfo {year} {1996})}\BibitemShut
  {NoStop}%
\bibitem [{\citenamefont {Kim}\ and\ \citenamefont
  {Mitani}(2005)}]{kim2005thermoelectric}%
  \BibitemOpen
  \bibfield  {author} {\bibinfo {author} {\bibfnamefont {D.-H.}\ \bibnamefont
  {Kim}}\ and\ \bibinfo {author} {\bibfnamefont {T.}~\bibnamefont {Mitani}},\
  }\bibfield  {title} {\bibinfo {title} {Thermoelectric properties of
  fine-grained bi2te3 alloys},\ }\href@noop {} {\bibfield  {journal} {\bibinfo
  {journal} {Journal of Alloys and Compounds}\ }\textbf {\bibinfo {volume}
  {399}},\ \bibinfo {pages} {14} (\bibinfo {year} {2005})}\BibitemShut
  {NoStop}%
\bibitem [{\citenamefont {Shi}\ \emph {et~al.}(2019)\citenamefont {Shi},
  \citenamefont {Sun}, \citenamefont {Bu}, \citenamefont {Zhang}, \citenamefont
  {Wu}, \citenamefont {Lin}, \citenamefont {Li}, \citenamefont {Faghaninia},
  \citenamefont {Jain},\ and\ \citenamefont {Pei}}]{shi2019revelation}%
  \BibitemOpen
  \bibfield  {author} {\bibinfo {author} {\bibfnamefont {X.}~\bibnamefont
  {Shi}}, \bibinfo {author} {\bibfnamefont {C.}~\bibnamefont {Sun}}, \bibinfo
  {author} {\bibfnamefont {Z.}~\bibnamefont {Bu}}, \bibinfo {author}
  {\bibfnamefont {X.}~\bibnamefont {Zhang}}, \bibinfo {author} {\bibfnamefont
  {Y.}~\bibnamefont {Wu}}, \bibinfo {author} {\bibfnamefont {S.}~\bibnamefont
  {Lin}}, \bibinfo {author} {\bibfnamefont {W.}~\bibnamefont {Li}}, \bibinfo
  {author} {\bibfnamefont {A.}~\bibnamefont {Faghaninia}}, \bibinfo {author}
  {\bibfnamefont {A.}~\bibnamefont {Jain}},\ and\ \bibinfo {author}
  {\bibfnamefont {Y.}~\bibnamefont {Pei}},\ }\bibfield  {title} {\bibinfo
  {title} {Revelation of inherently high mobility enables mg3sb2 as a
  sustainable alternative to n-bi2te3 thermoelectrics},\ }\href@noop {}
  {\bibfield  {journal} {\bibinfo  {journal} {Advanced Science}\ }\textbf
  {\bibinfo {volume} {6}},\ \bibinfo {pages} {1802286} (\bibinfo {year}
  {2019})}\BibitemShut {NoStop}%
\bibitem [{\citenamefont {Zhang}\ \emph {et~al.}(2011)\citenamefont {Zhang},
  \citenamefont {Yu}, \citenamefont {Feng}, \citenamefont {Yao}, \citenamefont
  {Weng}, \citenamefont {Dai},\ and\ \citenamefont
  {Fang}}]{zhang2011topological}%
  \BibitemOpen
  \bibfield  {author} {\bibinfo {author} {\bibfnamefont {W.}~\bibnamefont
  {Zhang}}, \bibinfo {author} {\bibfnamefont {R.}~\bibnamefont {Yu}}, \bibinfo
  {author} {\bibfnamefont {W.}~\bibnamefont {Feng}}, \bibinfo {author}
  {\bibfnamefont {Y.}~\bibnamefont {Yao}}, \bibinfo {author} {\bibfnamefont
  {H.}~\bibnamefont {Weng}}, \bibinfo {author} {\bibfnamefont {X.}~\bibnamefont
  {Dai}},\ and\ \bibinfo {author} {\bibfnamefont {Z.}~\bibnamefont {Fang}},\
  }\bibfield  {title} {\bibinfo {title} {Topological aspect and quantum
  magnetoresistance of $\beta$- ag 2 te},\ }\href@noop {} {\bibfield  {journal}
  {\bibinfo  {journal} {Physical review letters}\ }\textbf {\bibinfo {volume}
  {106}},\ \bibinfo {pages} {156808} (\bibinfo {year} {2011})}\BibitemShut
  {NoStop}%
\bibitem [{\citenamefont {Ge}\ \emph {et~al.}(2016)\citenamefont {Ge},
  \citenamefont {Liu}, \citenamefont {Feng}, \citenamefont {Lin},\ and\
  \citenamefont {He}}]{ge2016high}%
  \BibitemOpen
  \bibfield  {author} {\bibinfo {author} {\bibfnamefont {Z.-H.}\ \bibnamefont
  {Ge}}, \bibinfo {author} {\bibfnamefont {X.}~\bibnamefont {Liu}}, \bibinfo
  {author} {\bibfnamefont {D.}~\bibnamefont {Feng}}, \bibinfo {author}
  {\bibfnamefont {J.}~\bibnamefont {Lin}},\ and\ \bibinfo {author}
  {\bibfnamefont {J.}~\bibnamefont {He}},\ }\bibfield  {title} {\bibinfo
  {title} {High-performance thermoelectricity in nanostructured earth-abundant
  copper sulfides bulk materials},\ }\href@noop {} {\bibfield  {journal}
  {\bibinfo  {journal} {Advanced Energy Materials}\ }\textbf {\bibinfo {volume}
  {6}},\ \bibinfo {pages} {1600607} (\bibinfo {year} {2016})}\BibitemShut
  {NoStop}%
\bibitem [{\citenamefont {Wu}\ \emph {et~al.}(2018)\citenamefont {Wu},
  \citenamefont {Zhou},\ and\ \citenamefont {Hu}}]{wu2018two}%
  \BibitemOpen
  \bibfield  {author} {\bibinfo {author} {\bibfnamefont {B.}~\bibnamefont
  {Wu}}, \bibinfo {author} {\bibfnamefont {Y.}~\bibnamefont {Zhou}},\ and\
  \bibinfo {author} {\bibfnamefont {M.}~\bibnamefont {Hu}},\ }\bibfield
  {title} {\bibinfo {title} {Two-channel thermal transport in
  ordered--disordered superionic ag2te and its traditionally contradictory
  enhancement by nanotwin boundary},\ }\href@noop {} {\bibfield  {journal}
  {\bibinfo  {journal} {The Journal of Physical Chemistry Letters}\ }\textbf
  {\bibinfo {volume} {9}},\ \bibinfo {pages} {5704} (\bibinfo {year}
  {2018})}\BibitemShut {NoStop}%
\bibitem [{\citenamefont {Madsen}\ \emph {et~al.}(2016)\citenamefont {Madsen},
  \citenamefont {Katre},\ and\ \citenamefont {Bera}}]{madsen2016calculating}%
  \BibitemOpen
  \bibfield  {author} {\bibinfo {author} {\bibfnamefont {G.~K.}\ \bibnamefont
  {Madsen}}, \bibinfo {author} {\bibfnamefont {A.}~\bibnamefont {Katre}},\ and\
  \bibinfo {author} {\bibfnamefont {C.}~\bibnamefont {Bera}},\ }\bibfield
  {title} {\bibinfo {title} {Calculating the thermal conductivity of the
  silicon clathrates using the quasi-harmonic approximation},\ }\href@noop {}
  {\bibfield  {journal} {\bibinfo  {journal} {physica status solidi (a)}\
  }\textbf {\bibinfo {volume} {213}},\ \bibinfo {pages} {802} (\bibinfo {year}
  {2016})}\BibitemShut {NoStop}%
\bibitem [{\citenamefont {Katre}\ \emph {et~al.}(2017)\citenamefont {Katre},
  \citenamefont {Carrete}, \citenamefont {Dongre}, \citenamefont {Madsen},\
  and\ \citenamefont {Mingo}}]{katre2017exceptionally}%
  \BibitemOpen
  \bibfield  {author} {\bibinfo {author} {\bibfnamefont {A.}~\bibnamefont
  {Katre}}, \bibinfo {author} {\bibfnamefont {J.}~\bibnamefont {Carrete}},
  \bibinfo {author} {\bibfnamefont {B.}~\bibnamefont {Dongre}}, \bibinfo
  {author} {\bibfnamefont {G.~K.}\ \bibnamefont {Madsen}},\ and\ \bibinfo
  {author} {\bibfnamefont {N.}~\bibnamefont {Mingo}},\ }\bibfield  {title}
  {\bibinfo {title} {Exceptionally strong phonon scattering by b substitution
  in cubic sic},\ }\href@noop {} {\bibfield  {journal} {\bibinfo  {journal}
  {Physical review letters}\ }\textbf {\bibinfo {volume} {119}},\ \bibinfo
  {pages} {075902} (\bibinfo {year} {2017})}\BibitemShut {NoStop}%
\bibitem [{\citenamefont {Giannozzi}\ \emph {et~al.}(2009)\citenamefont
  {Giannozzi}, \citenamefont {Baroni}, \citenamefont {Bonini}, \citenamefont
  {Calandra}, \citenamefont {Car}, \citenamefont {Cavazzoni}, \citenamefont
  {Ceresoli}, \citenamefont {Chiarotti}, \citenamefont {Cococcioni},
  \citenamefont {Dabo} \emph {et~al.}}]{giannozzi2009quantum}%
  \BibitemOpen
  \bibfield  {author} {\bibinfo {author} {\bibfnamefont {P.}~\bibnamefont
  {Giannozzi}}, \bibinfo {author} {\bibfnamefont {S.}~\bibnamefont {Baroni}},
  \bibinfo {author} {\bibfnamefont {N.}~\bibnamefont {Bonini}}, \bibinfo
  {author} {\bibfnamefont {M.}~\bibnamefont {Calandra}}, \bibinfo {author}
  {\bibfnamefont {R.}~\bibnamefont {Car}}, \bibinfo {author} {\bibfnamefont
  {C.}~\bibnamefont {Cavazzoni}}, \bibinfo {author} {\bibfnamefont
  {D.}~\bibnamefont {Ceresoli}}, \bibinfo {author} {\bibfnamefont {G.~L.}\
  \bibnamefont {Chiarotti}}, \bibinfo {author} {\bibfnamefont {M.}~\bibnamefont
  {Cococcioni}}, \bibinfo {author} {\bibfnamefont {I.}~\bibnamefont {Dabo}},
  \emph {et~al.},\ }\bibfield  {title} {\bibinfo {title} {Quantum espresso: a
  modular and open-source software project for quantum simulations of
  materials},\ }\href@noop {} {\bibfield  {journal} {\bibinfo  {journal}
  {Journal of physics: Condensed matter}\ }\textbf {\bibinfo {volume} {21}},\
  \bibinfo {pages} {395502} (\bibinfo {year} {2009})}\BibitemShut {NoStop}%
\bibitem [{\citenamefont {Perdew}\ \emph {et~al.}(1996)\citenamefont {Perdew},
  \citenamefont {Burke},\ and\ \citenamefont
  {Ernzerhof}}]{perdew1996generalized}%
  \BibitemOpen
  \bibfield  {author} {\bibinfo {author} {\bibfnamefont {J.~P.}\ \bibnamefont
  {Perdew}}, \bibinfo {author} {\bibfnamefont {K.}~\bibnamefont {Burke}},\ and\
  \bibinfo {author} {\bibfnamefont {M.}~\bibnamefont {Ernzerhof}},\ }\bibfield
  {title} {\bibinfo {title} {Generalized gradient approximation made simple},\
  }\href@noop {} {\bibfield  {journal} {\bibinfo  {journal} {Physical review
  letters}\ }\textbf {\bibinfo {volume} {77}},\ \bibinfo {pages} {3865}
  (\bibinfo {year} {1996})}\BibitemShut {NoStop}%
\bibitem [{\citenamefont {Togo}\ and\ \citenamefont
  {Tanaka}(2015)}]{togo2015first}%
  \BibitemOpen
  \bibfield  {author} {\bibinfo {author} {\bibfnamefont {A.}~\bibnamefont
  {Togo}}\ and\ \bibinfo {author} {\bibfnamefont {I.}~\bibnamefont {Tanaka}},\
  }\bibfield  {title} {\bibinfo {title} {First principles phonon calculations
  in materials science},\ }\href@noop {} {\bibfield  {journal} {\bibinfo
  {journal} {Scripta Materialia}\ }\textbf {\bibinfo {volume} {108}},\ \bibinfo
  {pages} {1} (\bibinfo {year} {2015})}\BibitemShut {NoStop}%
\end{thebibliography}%

\end{document}